\documentclass[twocolumn]{aastex631}
\usepackage{graphicx} 

\usepackage{amsmath}
\usepackage{natbib}
\usepackage{array}
\usepackage{tabularx}
\usepackage{multirow}
\usepackage{makecell}

\begin{document}

\title{Impacts of UV Radiation from an AGN on Planetary Atmospheres and Consequences for Galactic Habitability}

\correspondingauthor{Kendall I. Sippy}
\email{ksippy@middlebury.edu}

\author[0009-0009-1099-7135]{Kendall I. Sippy} 
\affil{Department of Physics, Middlebury College, Middlebury, VT 05753, USA}
\affil{Department of Physics and Astronomy, Dartmouth College, 6127 Wilder Laboratory, Hanover, NH 03755, USA}
\author[0000-0001-5460-8159]{Jake K. Eager-Nash}
\affiliation{School of Earth and Ocean Sciences, University of Victoria, Victoria, BC, V8P 5C2, Canada}
\affiliation{Department of Physics and Astronomy, University of Exeter, Stocker Road, Exeter, EX4 4QL, UK}
\author[0000-0003-1468-9526]{Ryan C. Hickox}
\affil{Department of Physics and Astronomy, Dartmouth College, 6127 Wilder Laboratory, Hanover, NH 03755, USA}
\author[{0000-0001-6707-4563}]{Nathan J. Mayne}
\affiliation{Department of Physics and Astronomy, University of Exeter, Stocker Road, Exeter, EX4 4QL, UK}
\author[0000-0002-4024-6967]{McKinley C. Brumback}
\affil{Department of Physics, Middlebury College, Middlebury, VT 05753, USA}


\shortauthors{Sippy et. al}
\date{January 2025}

\begin{abstract}
    We present a study of the effects of ultraviolet (UV) emission from active galactic nuclei (AGN) on the atmospheric composition of planets and potential impact on life. It is expected that all supermassive black holes, which reside at galactic centers, have gone through periods of high AGN activity in order to reach their current masses. We examine potential damaging effects on lifeforms on planets with different atmosphere types and receiving different levels of AGN flux, using data on the sensitivity of various species' cells to UV radiation to determine when radiation becomes ``dangerous''. We also consider potential chemical changes to planetary atmospheres as a result of UV radiation from AGN, using the \texttt{PALEO} photochemical model. We find the presence of sufficient initial oxygen (surface mixing ratio $\geq 10^{-3} \rm\, mol/mol$) in the planet's atmosphere allows a thicker ozone layer to form in response to AGN radiation, which reduces the level of dangerous UV radiation incident on the planetary surface from what it was in absence of an AGN. We estimate the fraction of solar systems in galaxies that would be affected by substantial AGN UV radiation, and find that the impact is most pronounced in compact galaxies such as ``red nugget relics'', as compared to typical present-day ellipticals and spirals (using M87 and the Milky Way as examples).
\end{abstract}


\section{Introduction}
\label{sec:intro}

Supermassive black holes (SMBHs) exist at the center of virtually every massive galaxy and are known to grow via periods of intense accretion activity \citep{hopkins2006, volonteri2010}. There is evidence that our own galaxy's central black hole (BH), Sagittarius A*, experienced an active galactic nucleus (AGN) phase a few million years ago, as suggested by the observed Fermi bubbles \citep{su2010, zhang2020} as well as the unexpectedly high population of young stars near Sagittarius A* \citep{chen2015}.

Previous works have considered the risks to planetary habitability due to radiation from an AGN, primarily considering the dangers posed by Sagittarius A* in an active phase, from direct radiation \citep[e.g.,][]{Balbi2017,  Amar2019, BBB2019, Pac2020} as well as from atmospheric mass loss \citep[e.g.,][]{Balbi2017, Chen2018, Forbes2018, Wis2019,Ishi2024}. Of these, to our knowledge only \citet{BBB2019} consider risks stemming from UV radiation of an AGN specifically, using a framework where they argue that there may be danger to life on the planetary surface if the top-of-atmosphere (TOA) flux received from the AGN is equal to or greater than the total TOA flux of the Sun on Earth--- in other words if the TOA flux received on a hypothetical Earthlike planet orbiting a Sunlike star is doubled. \citet{Pac2020} also use a similar framework to estimate UV radiation danger associated with tidal disruption events (TDEs) from a central SMBH. 

Our study on UV radiation from AGN builds upon this previous work by making use of the range of studies considering habitability of planets around M dwarf stars, which are known to be very active and expose their planets to high levels of ultraviolet (UV) and extreme-UV (XUV) fluence during flares \citep{Segur2010, Chen2021, Ridgway2023}. We modified the Platform for Atmosphere, Land, Earth, and Ocean model, \texttt{PALEO}, which had previously been used to study the effects of different stellar spectra on planetary atmospheres \citep{Jake2024}, to address the problem of high UV flux from an AGN.

\subsection{UV Radiation and Life} \label{sec:uv+life}

There are a number of effects that high levels of UV radiation, such as those produced in an M dwarf flare or AGN radiation, may have on planetary habitability. Depending on a planet's atmospheric composition, these effects can be helpful or harmful. It is possible that high UV radiation could severely hinder the development of life on a planet with an Earthlike atmosphere by preventing the formation of complex chemicals \citep{Balbi2017}, however, in lower doses the formation of complex chemicals could be triggered by UV radiation, which might aid the development of life \citep{Ranj2017, Rim2018, BBB2019}. There is also the possibility that UV radiation would have virtually no effect at all, if the atmosphere is composed such that it effectively blocks out UV light from reaching the surface \citep{OMal2019, Estrela2020}.

Exoplanets in the super-Earth/miniature Neptune regime could also be made more habitable as XUV radiation could trigger hydrodynamic escape of hydrogen-rich atmospheres \citep{Luger2015, Chen2018, Wis2019}. Simulations have shown that oxygen in an atmosphere can be resistant to hydrodynamic escape even while lighter elements like hydrogen escape \citep{Wis2019}, and that the presence of $\rm CO_2$ can potentially make the atmosphere as a whole resistant to hydrodynamic escape from XUV radiation \citep{Tian2009}. Therefore, depending on the composition of a planetary atmosphere, the risks of atmospheric escape might also be mitigated. It has also been demonstrated that ozone depletion of planetary atmospheres via ionizing radiation from M dwarf flares can be disastrous for life on the planet \citep{Thom2015}, as ozone is largely responsible for the protection of the planetary surface from UV radiation which is harmful to life \citep{Gebau2017}. The generation or depletion of ozone is in turn connected to the abundance of oxygen in the atmosphere.

It is clear that the atmospheric composition of a planet is a particularly important factor regarding how its overall habitability responds to high doses of radiation. Aside from the loss of atmospheric gas, there is also the potential for high UV flux to generate chemical reactions within the planetary atmosphere that alter its properties, as is seen for 3D models of terrestrial planets orbiting flaring M dwarf stars \citep{Chen2021, Ridgway2023}. In \citet{Ridgway2023}, the UV flux from the M dwarf flares is shown to generate a thicker ozone layer which persists and offers some protection from future flares. While we model Earthlike planets orbiting Sunlike stars for the purposes of this work, we motivate our research questions by drawing upon the effects that UV radiation (from M dwarf flares) has already been demonstrated to have on planetary atmospheres and therefore habitability.

Different biological species on Earth are known to have varying tolerances to UV radiation. In our analysis of effects that AGN radiation has on planetary habitability, we consider the levels of UV flux incident on the planetary surface and how those compare to the tolerances of various species, as well as the chemical changes AGN flux can cause in the planet's atmosphere and how the subsequent evolution of the atmosphere may improve or worsen conditions for species on the planetary surface.   

\subsection{Galaxies of Interest} \label{sec:galaxies}

In order to relate our study to the planetary populations of real galaxies, we consider examples of elliptical and spiral galaxies (M87 and the Milky Way [MW], respectively), as well as ``red nugget relic'' galaxies for which the impact of AGN radiation may be particularly pronounced. We use the stellar population distribution data from each galaxy as a proxy for the distribution of Earth-like planets orbiting Sun-like stars throughout the galaxies, to determine what percentage of these planets in a given galaxy will be affected by certain levels of AGN UV radiation. This stellar population analysis is described in more detail in \ref{sec:stellarpop}.

M87 is one of the most well-observed elliptical galaxies, containing the BH recently imaged with the Event Horizon Telescope at its center--- an SMBH of mass $5.9 \times 10^9$ $M_\odot$ \citep{Pri2016}. Due to its nature as a large elliptical, the potential planetary systems in this galaxy are far more susceptible to damage from its central SMBH, M87*. This galaxy has an almost complete lack of cold interstellar medium (ISM) which might attenuate radiation, a very massive central SMBH, and a compact stellar population distribution, with large portions of the stars near to M87's galactic center \citep{fan2008, Pri2016, whit2020}.   

We also consider the MW, as an example spiral galaxy and because these results have more direct impacts for humans and the search for extraterrestrial life in our neighborhood. Our central SMBH, Sgr A* (with a mass of $4.1 \times 10^6 M_\odot$) is known to have experienced active phases in the past, the most recent of which is estimated to have occurred a few million years ago \citep[e.g.][]{mezger1996, su2010, volonteri2010, bland2013, chen2015, zhang2020}.    

Finally, we consider the local relic galaxies Mrk 1216, NGC 1271, NGC 1277, NGC 384, UGC 2698, and PGC 11179. These relics, also called `compact elliptical galaxies' (CEGs) or ``red nuggets'', are local analogs for the expected progenitors of modern massive elliptical galaxies \citep{VanD2010, Yil2017}; these six in particular are selected because they have both stellar mass distributions and BH mass estimates in the literature. These galaxies all host SMBHs, with masses as follows: NGC 1271 = $3.0 \times 10^9 M_\odot$ \citep{Walsh2015}; NGC 1277 = $4.9 \times 10^9 M_\odot$ \citep{Walsh2016}; Mrk 1216 = $4.9 \times 10^9 M_\odot$ \citep{Walsh2017}; UGC 2698 = $2.46 \times 10^9 M_\odot$ \citep{Cohn2021}; PGC 11179 = $1.91 \times 10^9 M_\odot$ \citep{Cohn2023}; and NGC 384 = $4.34 \times 10^8 M_\odot$ \citep{Cohn2024}. This sample of relic galaxies allows us to probe how the planetary population of galaxies in the early universe might have been affected by UV radiation from AGN. Also, by virtue of having more compact stellar populations than large ellipticals like M87, they are good candidates to potentially have a large percentage of their stellar population be substantially affected by AGN radiation as opposed to typically more extended modern galaxies.

\section{Methods}
\label{sec:methods}
In this section, we present the methods by which we have analyzed the effect of AGN radiation on planetary atmospheres and biological species residing within them. In Section \ref{sec:AGNSEDs}, we describe how we scaled an AGN spectral energy distribution (SED) to represent various physical situations we aim to simulate. In Section \ref{sec:stellarpop}, we discuss how we estimate the percent of the stellar population in a given galaxy affected by different levels of AGN radiation. In Section \ref{sec:PALEO} we elaborate on the properties of the Platform for Atmosphere, Land, Earth, and Ocean model, \texttt{PALEO}, and configurations used; in particular our initial and boundary conditions for the atmosphere in \texttt{PALEO}, which represent surface oxygen mixing ratios typical of the Archean, Proterozoic, and Modern atmospheric compositions of the Earth. Further details on \texttt{PALEO} can be found in \citet{Jake2024}. Lastly, in Section \ref{sec:UVspecies}, we detail how we quantify danger to different species (humans, rats, \textit{E. Coli}, and \textit{D. Radiodurans}) due to UV radiation.\\

\subsection{AGN Radiation} \label{sec:AGNSEDs}
We use a mean quasar SED from \citet{Rich2006} as a model for the SED of an AGN, and consider the hypothetical conditions in our galaxies of interest if their central SMBHs were to enter an AGN phase. For simplicity we assume the AGN SED does not change with luminosity or Eddington ratio. We scale the mean quasar SED by factors such that their bolometric flux at the position of a hypothetical planet is equal to $10 \rm \,erg \,s^{-1} \, cm^{-2}$, $100 \rm \,erg \,s^{-1} \, cm^{-2}$, etc. up to $10^7 \rm \,erg \,s^{-1} \, cm^{-2}$. This scaling is chosen to cover the range of bolometric fluxes roughly associated with danger to different species on the surface of an Earthlike planet, as defined further in Section \ref{sec:UVspecies}. Table \ref{tab:scaling} provides the legend of what physical conditions (BH mass, Eddington ratio, and distance) each scaled flux could represent in the main three galaxies we discuss: M87, the MW, and NGC 1277. NGC 1277 is chosen as an example of the conditions in our red nugget relic galaxy sample, because it has the most dramatic result in terms of large percentages of the planets in the galaxy potentially receiving substantial doses of radiation (due to the combination of its BH mass and stellar population distribution; see Section \ref{sec:agnresults}), so it represents an upper limit on the impacts of AGN UV radiation on a galaxy's habitability. However, the results in the other red nuggets are qualitatively similar. A visual representation of this information is shown in Figure \ref{fig:straightlines} in the Appendix. The corresponding information for the other five red nuggets is provided in the Appendix, Table \ref{tab:scaling_app} in Section \ref{sec:rednuggets}.

\begin{table*}[t]
    \centering
    \begin{tabular}{cc||c|c|c|c|c|c|c|} \hline
    \multicolumn{2}{c}{\small Flux ($\rm erg\,s^{-1}\,cm^{-2}$) =}& $10$ & $10^2$ & $10^3$ & $10^4$ & $10^5$ & $10^6$ & $10^7$ \\ \hline
    \textbf{\footnotesize Galaxy} & \textbf{\footnotesize Edd. Ratio} & \multicolumn{7}{c|}{\textbf{\footnotesize Distance (pc)}} \\ \Xhline{2\arrayrulewidth}
    
    \multirow{3}{4em}{M87} & $1$ & $2.49 \times 10^{4}$ & $7.87 \times 10^{3}$ & $2.49 \times 10^{3}$ & $787$ & $249$ & $78.7$ & $24.9$\\
    & $0.1$ & $7.87 \times 10^{3}$ & $2.49 \times 10^{3}$ & $787$ & $249$ & $78.7$ & $24.9$ & $7.87$\\
    & $0.01$ & $2.49 \times 10^{3}$ & $787$ & $249$ & $78.7$ & $24.9$ & $7.87$ & $2.49$ \\ \hline

   \multirow{3}{4em}{MW Bulge} & $1$ & $656$ & $207$ & $65.6$ & $20.7$ & $6.56$ & $2.07$ & $0.656$\\
    & $0.1$ & $207$ & $65.6$ & $20.7$ & $6.56$ & $2.07$ & $0.656$ & $0.207$ \\ 
    & $0.01$ & $65.6$ & $20.7$ & $6.56$ & $2.07$ & $0.656$ & $0.207$ & $0.0656$\\ \hline

    \multirow{3}{4em}{NGC 1277} & $1$ & $2.27 \times 10^4$ & $7.17 \times 10^3$ & $2.27 \times 10^3$ & $717$ & $227$ & $71.7$ & $22.7$ \\
    & $0.1$ & $7.17 \times 10^3$ & $2.27 \times 10^3$ & $717$ & $227$ & $71.7$ & $22.7$ & $7.17$ \\
    & $0.01$ & $2.27 \times 10^3$ & $717$ & $227$ & $71.7$ & $22.7$ & $7.17$ & $2.27$ \\ \hline

    \end{tabular}
    \caption{\centering Correspondence of scaled total AGN fluxes to physical situations for different galaxies. By reading down the column of a certain flux in $\rm erg\,s^{-1}\,cm^{-2}$, we can read off all the Eddington ratio and distance (in $\rm pc$) combinations for a given galaxy that would be equivalent to the curves represented by this particular total AGN flux. Notably, since the flux values scale up and down by $100$, and the Eddington ratios scale the flux up or down by $10$, the distances for a given galaxy are all the same number scaled up or down by factors of $\sqrt{10}$. The initial number being scaled depends only on the black hole mass.}
    \label{tab:scaling}
\end{table*}

\subsection{Stellar Distributions} \label{sec:stellarpop}
We use the stellar mass distribution in our galaxies of interest to estimate what percentage of the galaxy's total stars would be receiving flux above certain limits that we define as dangerous to life in Section \ref{sec:UVspecies}. This is a proxy for estimating what percentage of possible habitable worlds in the galaxy would be affected by this radiation, assuming a spherically symmetric distribution of stars and a constant stellar mass function and mass-to-light ratio.

For the M87 galaxy, we use the radial stellar luminosity volume-density profile provided in Figure 1 of \citet{Geb2009}, and integrate in order to find a count of stars. For the MW, we only consider stars within the bulge of the galaxy. This region has a `boxy-peanut' shape, but for the sake of simplicity we also assume spherical symmetry. We use the stellar mass versus radius data for the MW bulge from Figure 2 Panel 3 of \citet{NewMWSofue2009}. The choice to only consider the central bulge of the MW is motivated by how limited the radiation received on hypothetical planets even within the bulge is, assuming no attenuation from the ISM (see Table \ref{tab:scaling}). In the disk, which has a much more substantial ISM and stars located much further from the galactic center \citep{Wyse1997}, the radiation is assumed to be negligible.

Lastly, for the red nugget relics, we use the surface brightness density versus projected radius data provided in Figure 6 of \citet{Yil2017}, and deproject and integrate the data to get a count of stars as a function of radial distance. The surface brightness profiles given in \citet{Yil2017} only extend down to $100 \rm\, pc$ from the galactic center at its minimum, but the stellar population at radii nearer to the center than this become important later on in our analysis. Therefore, we extrapolate the surface brightness profiles, $I(R)$, down to $R =10 \rm \, pc$ from the center--- this is done before any deprojection or integration \citep{Numpy}. We produce these extrapolated data points using a power law fit to the inner region of the surface brightness profile for each galaxy (projected radii  from $100$ to $300 \rm \, pc$; the choice of $300 \rm \, pc$ as the outer limit is arbitrary).

\subsection{Planetary Atmospheres} \label{sec:PALEO}
We model the effects that AGN UV radiation has on planetary atmospheres, which in turn affects habitability of the planetary surface, using the the Platform for Atmosphere, Land, Earth, and Ocean model, \texttt{PALEO}. \texttt{PALEO} was developed as a flexible framework for modelling the Earth System \citep[e.g.][]{Daines2016}, and more recently, the addition of the 1D photochemical atmosphere component has allowed \texttt{PALEO} to model the atmospheric chemistry of exoplanets \citep{Jake2024}. We used a configuration of \texttt{PALEO} with a 1D photochemical atmosphere, similar to that described in \citet{Jake2024}, to test the effects of AGN radiation on a planetary atmosphere.

Previously, \citet{Jake2024} used \texttt{PALEO} to investigate the evolution of atmospheres under radiation from a single source, comparing the effects of radiation from the Sun at $3.8 \rm\, Gyr$ ago versus the M dwarf Trappist 1. For our work, we have developed the \texttt{PALEO} framework to model photochemistry resulting from multiple irradiating bodies, allowing us to combine the SEDs of the host star and an AGN. Specifically, \texttt{PALEO} uses an SED in the UV, visible and near-infrared wavelengths, from approximately $117 \rm\, nm$ to $995 \rm \, nm$. This is the wavelength range that is relevant to photochemistry in a planetary atmosphere containing carbon, hydrogen, oxygen, sulfur, and nitrogen compounds. Our two inputted SEDs are that of the modern Sun, for simplicity's sake, and the AGN SED described in Section \ref{sec:AGNSEDs}. 

Notably, the AGN SED we use includes a wider range of wavelengths than those that \texttt{PALEO} requires, because \texttt{PALEO} only captures those wavelengths of light relevant to photochemistry of carbon, hydrogen, oxygen, nitrogen, and sulfur molecules. As such the bolometric fluxes ($F_{Bol}$) by which our simulations are labeled in all plots in Section \ref{sec:results} are bolometric fluxes integrated over the full wavelength range our SED covers, from approximately $64\rm\, nm$ to $0.142 \rm\, mm$. In practice, the total integrated flux over the range which \texttt{PALEO} considers is roughly $10\%$ of the AGN's bolometric flux as we define it. We have chosen this labeling method because using a value closer to the true bolometric luminosity of the AGN is more intuitive, and can then also be related directly to the Eddington ratio of the black hole (BH).

The atmosphere modeled in \texttt{PALEO} is cloud-free with a model top of $100 \rm\, km$ (divided into $200$ equally spaced levels in which reactions are computed), and assumes a constant flux at the top of the single column atmosphere for the entire model runtime, evenly distributed on all sides of the planet. We use the same parameterizations for the atmosphere described in \citet{Jake2024}, but we do not include the ocean component. We use the  ``full'' chemical network described in \citet{Jake2024}, which includes C, N, H, O, and S containing gaseous molecules. The full network of reactions between these can be found in Appendix A of \citet{Jake2024}. The boundary conditions and other initial parameters we use can be found in Table \ref{tab:paleoboundaries}.

\begin{table}[h]
    \centering
    \begin{tabular}{|c|c|} 
    \hline
    \textbf{\footnotesize Molecule} & \textbf{\footnotesize Surface mixing ratio ($\rm mol/mol$)} \\ \Xhline{2\arrayrulewidth}

    $\rm O_2$ & $0.21$, $1.0 \times 10^{-3}$, $1.0 \times 10^{-7}$ \\
    $\rm CO_2$ & $280.0 \times 10^{-6}$\\ 
    $\rm CH_4$ & $722.0 \times 10^{-9}$ \\ \hline

     & \textbf{\footnotesize Surface flux ($\rm cm^{-2} \, s^{-1}$)} \\ \Xhline{2\arrayrulewidth}

     $\rm CO$ & $3.7 \times 10^{11}$ \\ 
     $\rm C_2 H_6$ & $9.0 \times 10^{8}$ \\ 
     $\rm NO$ & $1.0 \times 10^{9}$ \\ 
     $\rm H_2 S$ & $2.0 \times 10^{8}$ \\ 
     $\rm OCS$ & $1.5 \times 10^{7}$ \\ 
     $\rm N_2 O$ & $1.53 \times 10^{9}$ \\ 
     $\rm SO_2$ & $9.0 \times 10^{9}$ \\ 
     $\rm H_2 SO_4$ & $7.0 \times 10^{8}$\\ \hline
    
    \end{tabular}

    \caption{\centering Surface boundary conditions set in \texttt{PALEO}, constant mixing ratios set at the surface of the planet and fixed surface fluxes, for relevant molecules. These values are consistent across our \texttt{PALEO} simulations, while the surface mixing ratio for molecular oxygen is varied between three values.}
    \label{tab:paleoboundaries}
\end{table}

For our AGN experiments, we use the chemical parameters to model the composition of Earth's atmosphere today, during the Archean period, and during the Proterozoic period by changing the surface $\rm O_2$ mixing ratio. We use these different epochs in Earth history as an attempt at accounting for different atmospheric compositions exoplanets may have, though we are still inherently very biased towards inhabited, terrestrial planets. The Archean period occurred on Earth from around $4 - 2.5 \rm \, Gyr$ ago, before the Great Oxidation Event, thus its very low $\rm O_2$ levels \citep{Bekk2004}. The Proterozoic period occurred directly after the Great Oxidation event, beginning roughly $2.4 \rm\, Gyr$ ago \citep{Bekk2004} and ending roughly $541 \rm\, Myr$ ago. The most important difference, for our analysis, between these three epochs in Earth's history is the oxygen mixing ratio, which is $0.21$ for the modern Earth, and we take values of $1\times 10^{-3}$ and $1\times 10^{-7}$ for the Proterozoic and Archean atmospheres, respectively. These represent two stable $\rm O_2$ surface mixing ratios on either side of the Great Oxidation Event \citep{Greg2021}, though the exact concentration of oxygen likely changed over these eras. We only vary the concentration of oxygen between our three model atmospheres (see Table \ref{tab:paleoboundaries}), as this the most significant difference between these three epochs of Earth's history. Furthermore, the most influential chemical process occurring in our simulations is the formation of ozone, $\rm O_3$, via the Chapman mechanism \citep{Chap1930}, which is driven by UV flux. The varying $\rm O_2$ levels in our atmospheric simulations dictate the amount of ozone that is ultimately able to form.

We run simulations for $5 \rm \,Myr$ to allow for an equilibrium to be reached, which we determine as when the mixing ratios of gasses are no longer evolving. We initialize our AGN runs using the equilibrium composition of the Sun-only simulations, to allow us to investigate the evolution of the atmosphere following the onset of an AGN phase. We use the SED of the modern Sun in all experiments, seeing as the difference in UV flux between the prehistoric Sun and the modern Sun is negligible compared to the UV flux provided by the AGN in our simulations \citep{Claire2012}. In the highest AGN flux simulation we ran under the Archean atmosphere, the tolerance of the solver was increased to allow the simulation to fully equilibrate. In addition to chemical concentration as a function of altitude, \texttt{PALEO} also outputs the integrated spectrum of transmitted flux of the atmosphere (for light passing from the top to the bottom of the atmosphere), which in turn gives us the surface flux on the model planet. We can compare the input SED, also called the TOA flux, to surface flux to assess how much or little protection from UV radiation the atmosphere provides.

\subsection{UV Tolerance of Species} \label{sec:UVspecies}

In order to quantify the level of UV flux at which organisms on a planetary surface may be endangered, we turn to several biological studies considering the average dose of UV radiation to kill $50\%$ of a sample of cells ($\rm LD_{50}$) from several lifeforms--- humans, rats, \textit{E. Coli}, and \textit{D. Radiodurans}--- as the ``limit of danger''. The flux limits we discuss do not imply that the species in question will face immediate death upon experiencing this level of flux, simply that there are dangers to individual organisms' health, and potentially to the species' long-term survival if this level of radiation persists for a significant period of time. In Section \ref{sec:finalresults}, it can be observed that even radiation from the Sun surpasses some of these danger limits. This is indeed true, in the sense that an organism experiencing 24 hours of continuous radiation from the Sun with no kind of shade or other sun protection has an increased risk of long term health damage \citep{dora2013}, and that if this persists for all members of the species for extended periods of time, there are risks to their long term survival.

We use data for human epidermal keratinocytes, the cell type that makes up the majority of the skin's outermost layer \citep{AokiHumans2011}, rat PC12 cells derived from rat pheochromocytoma tumors \citep{MasumaRats2013}, \textit{E. Coli}, a common bacteria, and \textit{D. Radiodurans}, an extremophile known to be very resistant to high radiation levels \citep{EColiDRad2010}. \citet{EColiDRad2010}, \citet{AokiHumans2011}, and \citet{MasumaRats2013} provide lethal doses to these cell types over either $24$ or $48 \rm\, hr$ of UV exposure, at a variety of wavelengths. We standardize them all to $24\rm\, hr$, assuming that the total UV fluence received is what defines the limit of danger. We refer to these data as ``spectra'' throughout, but note that they represent a differential flux incident on cells required to generate a certain effect, not a flux radiated from a body.  

The lethal UV doses to human keratinocytes provided in \citet{AokiHumans2011} cover a wavelength range from $235\rm\, nm$ to $310\rm\, nm$, whereas the doses for rats in \citet{MasumaRats2013} range from $250 \rm\, nm$ to $310 \rm\, nm$, and in \citet{EColiDRad2010} the lethal dose is only provided at $260 \rm\, nm$ for \textit{E. Coli} and \textit{D. Radiodurans}. Observing the similar shape of the spectra of fluxes lethal to $50\%$ of humans and rat cells, we extrapolate all of the spectra out to the wavelength limits of the human cell lethality spectrum. We perform this extrapolation by assuming each of the spectra share the same shape, scaled up or down by a constant factor. The original and extrapolated spectra are shown in the Appendix, Figure \ref{fig:species_both} in Section \ref{sec:extrap}, for reference. 

All three studies providing the lethal dose data to cells uses bulbs with a $10 \rm\, nm$ bandwidth to provide UV light at each wavelength. Using this, we can convert their flux data into a differential flux with respect to wavelength, $F_\lambda$, assuming the spectrum of each bulb has a `top hat' shape and delivers a constant flux at each wavelength in its bandwidth. We then convert to a differential flux with respect to frequency, $F_\nu$, which matches the units of the AGN SED and can be compared.  

\section{Results and Discussion}
\label{sec:results}
In this section, we give an overview of how UV radiation from an AGN affects planetary atmospheres and therefore habitability. First, in Section \ref{sec:atmevolution} we present the results of our simulations in \texttt{PALEO}, showing how different levels of AGN flux alter the atmospheric composition and transmission of UV radiation over time. We also make note of the possibility of a runaway greenhouse effect in some cases, which \texttt{PALEO} does not simulate. Then, in Section \ref{sec:finalresults}, we compare the UV transmission of the various model planetary atmospheres before and after their composition was changed by AGN radiation. We comment on how the danger to species on the planetary surface has evolved as a result of this. We estimate based on AGN radiation alone (excluding solar radiation) the percentage of the potential habitable worlds which may be in danger in our galaxies of interest, which were described in Section \ref{sec:galaxies}.

\subsection{Atmospheric Evolution with \texttt{PALEO}} \label{sec:atmevolution}
High levels of UV radiation from an AGN will not only be radiatively transported through an existing atmosphere, but can cause chemical reactions that alter atmospheric composition, as discussed in Section \ref{sec:uv+life}. Here we use the atmospheric modeling code \texttt{PALEO} (described in Section \ref{sec:PALEO}) to study these changes.

\subsubsection{Modern Earth Atmosphere} \label{sec:modatm}
\begin{figure}
    \centering
    \includegraphics[scale=0.55]{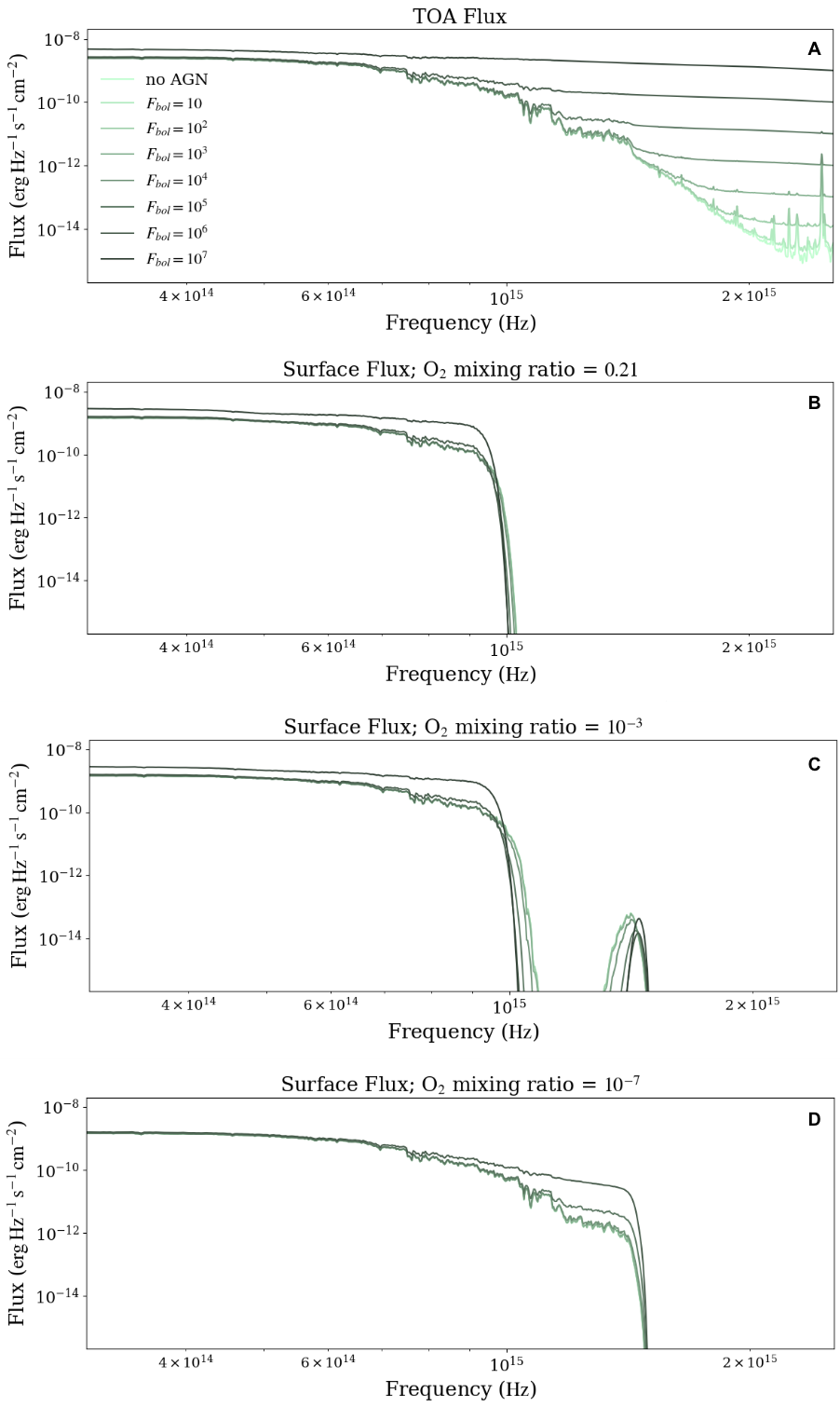}
    \caption{Flux in $\rm erg\,s^{-1}\,cm^{-2}\,Hz^{-1}$ vs. frequency in $\rm Hz$ received on a hypothetical planet with an modern Earth atmosphere for various AGN scaling factors, plus the flux from the Sun, in \texttt{PALEO}. Curves are labeled according to bolometric AGN flux in $\rm erg \, s^{-1}\, cm^{-2}$, per the legend in panel A. See Table \ref{tab:scaling} for physical situations corresponding to each level of AGN flux. Panel A: TOA flux, same across all simulations. Panel B: Surface flux under the Modern atmosphere ($\rm O_2$ mixing ratio $0.21 \rm\, mol/mol$). We notice a substantial drop-off in surface flux in the UV. Panel C: Surface flux under the Proterozoic atmosphere ($\rm O_2$ mixing ratio $10^{-3} \rm\, mol/mol$). This case is relatively similar to the Modern, though with more flux transmitted. Panel D: Surface flux under the Archean atmosphere ($\rm O_2$ mixing ratio $10^{-7} \rm\, mol/mol$). In this case, there is substantial UV surface flux. Note that there is no $F_{bol} = 10^7 \rm \, erg \, s^{-1} \, cm^{-2}$ simulation for the Archean atmosphere.}
    \label{fig:all_flux}
\end{figure}

\begin{figure*}
    \centering
    \includegraphics[scale=0.5]{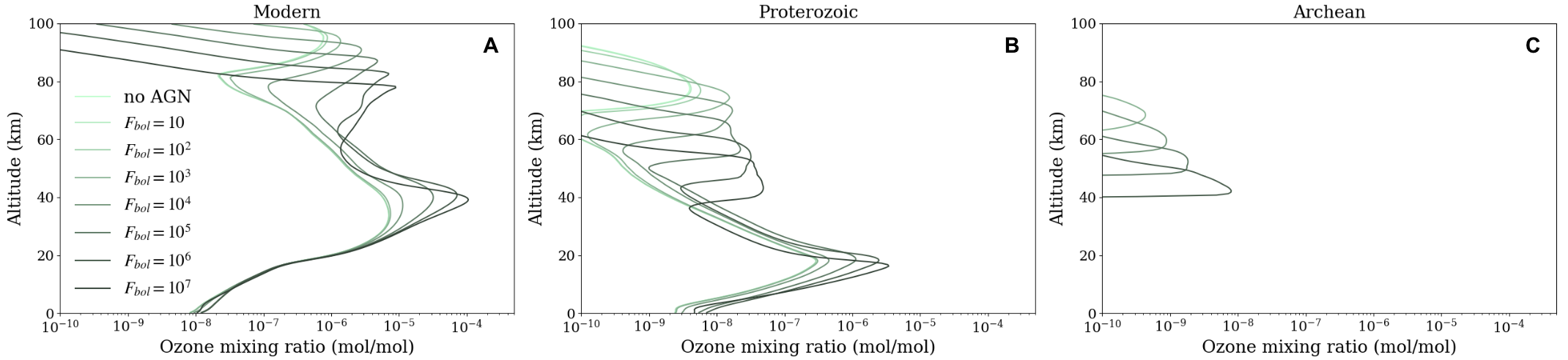}
    \caption{Ozone mixing ratio in $\rm mol/mol$ for the same simulations shown in Figure \ref{fig:all_flux} in \texttt{PALEO}, at different altitudes in $\rm km$ of the model atmosphere. Panel A: Modern atmosphere. Substantial ozone production occurs in higher flux simulations. Panel B: Proterozoic atmosphere. Substantial ozone production also occurs, though ultimately lower concentrations than the Modern are produced, and lower in the atmosphere. Panel C: Archean atmosphere. Very little ozone production occurs.}
    \label{fig:all_ozone}
\end{figure*}

First, we run the model for the Modern Earth atmosphere, with AGN SEDs scaled to various bolometric fluxes. The TOA flux applied to all simulations (which is consistent across all three atmosphere types) is shown in Figure \ref{fig:all_flux}A. The results for the Modern atmosphere are shown in Figure \ref{fig:all_flux}B and Figure \ref{fig:all_ozone}A.

In Figure \ref{fig:all_flux}A, we see that the main difference in the AGN radiation versus the No AGN case is the high UV radiation introduced. The AGN spectrum is much bluer than the solar spectrum, so the solar radiation dominates at lower frequencies, while the AGN radiation dominates around the UV. The surface spectrum nevertheless has a negligible flux transmitted through the atmosphere in the UV regardless of the AGN flux (see Figure \ref{fig:all_flux}B). The explanation for these results lies in Figure \ref{fig:all_ozone}A, where we see how the atmospheric composition of the model has evolved for the AGN cases. Based on these results we hypothesize that increased UV flux caused chemical reactions of $\rm O_2$ in the atmosphere into $\rm O_3$ via the Chapman mechanism, which then provided greater protection against UV radiation, as was the case in \citet{Ridgway2023} for a model planet receiving excess UV flux from a flaring M dwarf star.    

\subsubsection{Proterozoic Earth Atmosphere} \label{sec:protatm}
Next, we tested an intermediate case, the Proterozoic Earth atmosphere, which has a greater concentration of oxygen than the Archean, but less than the modern Earth. The results for these experiments are shown in Figure \ref{fig:all_flux}C, and Figure \ref{fig:all_ozone}B.

In this case, we see once again that there is attenuation in the UV at the planet's surface, though less so than the Modern Earth atmosphere. Interestingly, adding more flux to the simulation is actually reducing the amount of flux which is ultimately transmitted to the surface from $F_{bol} = 10^4 \rm\, erg\, s^{-1} \, cm^{-2}$ and upwards in AGN flux (darker green curves). This is also the case in the Modern atmosphere (Figure \ref{fig:all_flux}B), but the effect is much more pronounced for the Proterozoic. The Archean (Figure \ref{fig:all_flux}D, discussed in Section \ref{sec:archatm}) does not exhibit this behavior. This result suggests that the concentration of oxygen in the Proterozoic atmosphere represents an intermediate case, where the production of UV-protecting ozone is highly sensitive to small changes in the flux. The lower initial oxygen level means this rapid ozone protection produces a more substantial increase in surface protection from UV radiation as compared to the Modern or Archean atmospheres. 
In the ozone mixing ratio plot for these simulations, Figure \ref{fig:all_ozone}B, at $10^4 \rm \, erg\,s^{-1}\,cm^{-2}$ of AGN flux and above the ozone layer in the upper atmosphere becomes notably thicker and moves to a somewhat lower altitude. This is consistent with that we observed in the surface flux plot (Figure \ref{fig:all_flux}C).

\subsubsection{Archean Earth Atmosphere} \label{sec:archatm}
Since the formation of an ozone layer is dependent on sufficient molecular oxygen, we now examine $\rm O_2$ surface mixing ratios that are not sufficient to form an ozone layer, as occurred during the Archean period. The results for the Archean atmosphere are shown in Figure \ref{fig:all_flux}D and Figure \ref{fig:all_ozone}C. The $F_{bol} = 10^7 \rm \, erg \, s^{-1} \, cm^{-2}$ simulation for the Archean atmosphere is not included because it did not converge, due to the high level of flux combined with the very low level of oxygen.

In the surface flux plot, we see that there is now virtually no attenuation in the UV. With this atmosphere, even the No AGN/Sun-only case provides a high dose of UV radiation to the surface, though we note that simple, monocellular life did exist on Earth's surface during this time \citep{Lep2020}, so these conditions are, in principle, habitable. In terms of the ozone concentration in the Archean atmosphere simulations, Figure \ref{fig:all_ozone}C, we can see that there is minimal ozone production compared to the other two atmospheres, which is consistent with the limited amount of oxygen that existed initially. The fact that little ozone is produced also reaffirms the key role of ozone production to planetary surface protection from UV radiation in these simulations.

\subsubsection{Time Evolution} \label{sec:timeevo}
As discussed in Sections \ref{sec:modatm}, \ref{sec:protatm}, and \ref{sec:archatm}, the increased flux from the AGN leads to the increase in ozone mixing ratio in the atmosphere in the Modern and Proterozoic atmospheres (and not in the low-oxygen Archean atmosphere), which in turn protects the surface from potentially dangerous UV radiation. Next, we quantify how quickly this UV-protecting ozone layer develops, in order to determine how long the initial radiation danger level persists before it is mitigated by a thickening of the ozone layer.

\begin{figure*}
    \centering
    \includegraphics[scale=0.6]{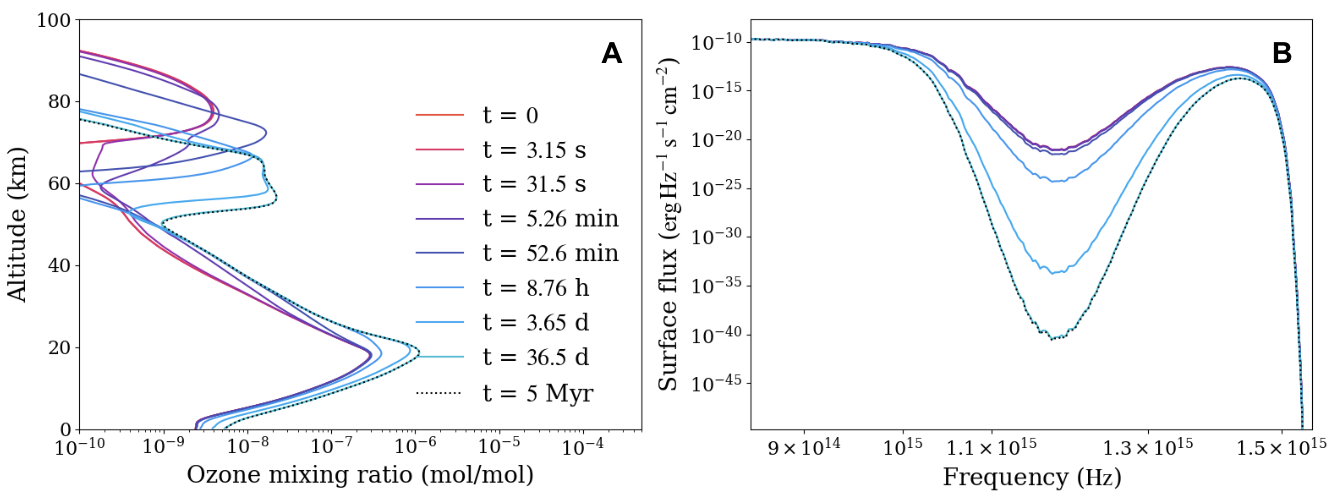}
    \caption{Time evolution of the ozone layer and surface radiation received in the $F_{bol} = 10^5 \rm \, erg \, s^{-1} \, cm^{-2}$ AGN simulation in \texttt{PALEO}, with a Proterozoic atmosphere. The different color curves represent times in the simulation as the atmosphere evolves. Panel A: Ozone mixing ratio in $\rm mol/mol$ in the model atmosphere versus altitude in $\rm km$. Panel B: Surface flux received in $\rm erg\, s^{-1}\, cm^{-2} \, Hz^{-1}$ versus frequency in $\rm Hz$. Only the final state ($t = 5 \rm\, Myr$, dotted black line on both panels) of the ozone mixing ratio and surface flux are shown after $t = 36.5 \rm\, days$, as there is virtually no change beyond that time.}
    \label{fig:time}
\end{figure*}

The ozone formation at different times for a Proterozoic Earth atmosphere with radiation scaled to $10^5 \rm \,erg\,s^{-1}\,cm^{-2}$ is shown in Figure \ref{fig:time}. We model this particular flux level in the Proterozoic atmosphere here because it represents a limiting case where small (order of magnitude) increases in TOA flux are generating substantial amounts of ozone production (see Section \ref{sec:protatm}). In the first approximately $1 \rm\, month$ of simulated time, there is a dramatic evolution of the ozone distribution from its initial atmospheric conditions having experienced only solar radiation (red curve, representing $t = 0$, Figure \ref{fig:time}A). We see that after $t = 36.5 \rm\, days$, the ozone distribution and surface flux virtually stop changing. The lack of change in the surface flux indicates the atmospheric transmission is no longer changing. These results are consistent with what was found for planetary atmospheres experienced UV radiation from M dwarf flares in \citet{Ridgway2023}, where after the initial simulated M dwarf flare the atmospheric ozone level rises to what will ultimately be its final level before $\sim 50 \rm \, days$, as shown in their Figure 9. 

Relative to the timescales we are considering for AGN activity, which are on the order of $10^5 \rm\, yr$ \citep{Schaw2015}, this ozone development happens virtually instantaneously. However, for the species living on the planetary surface, this is still a significant period of time to experience high irradiation. When we quantified danger to species (see Section \ref{sec:UVspecies}) we considered the fluence necessary to kill cells on a timescale of $24 \rm\, hr$. So while the timescale for atmospheric evolution is negligible compared to the time the AGN will be active, there will still be significant risks to species on the surface during this $\sim 1 \rm \, month$ period, assuming a constant AGN flux from the time the AGN ``turns on''.

\subsubsection{Runaway Greenhouse}\label{sec:runawaygreenhouse}
The \texttt{PALEO} atmospheric modeling code we use in this work does not model the evolving climate - only chemistry evolves under a fixed pressure-temperature profile. This means we cannot model the runaway greenhouse transition, which involves heating of the atmosphere. However, if the incoming TOA flux exceeds a certain threshold, we can predict this phenomenon will occur. The limit for a runaway greenhouse process to occur on a planet has been estimated at a TOA flux of $1500 \rm \, W/m^2$ in solar radiation \citep{Leconte2013}, equivalent to $1.5 \times 10^{6} \rm\, erg \,s^{-1} \,cm^{-2}$. To compare our models to this threshold, we integrate the AGN flux over the full wavelength range of our original AGN SED from \citet{Rich2006}, and add the total flux from the Sun--- $1360 \rm \, W/m^2$, or $1.36 \times 10^6 \rm\, erg \, s^{-1} \, cm^{-2}$. 

The integrated incoming TOA flux from the AGN plus the Sun exceeds this limit at a bolometric AGN flux of $F_{bol} = 10^{5.15} \rm\, erg \, s^{-1} \, cm^{-2}$. If a runaway greenhouse does develop, then this may negate any benefits to species living on a planet with increased ozone protection developed from UV radiation, as the high temperature of the planetary surface may render it uninhabitable regardless.

\subsection{Results Post-Atmospheric Evolution} \label{sec:finalresults}

\begin{figure*}
    \centering
    \includegraphics[scale=1]{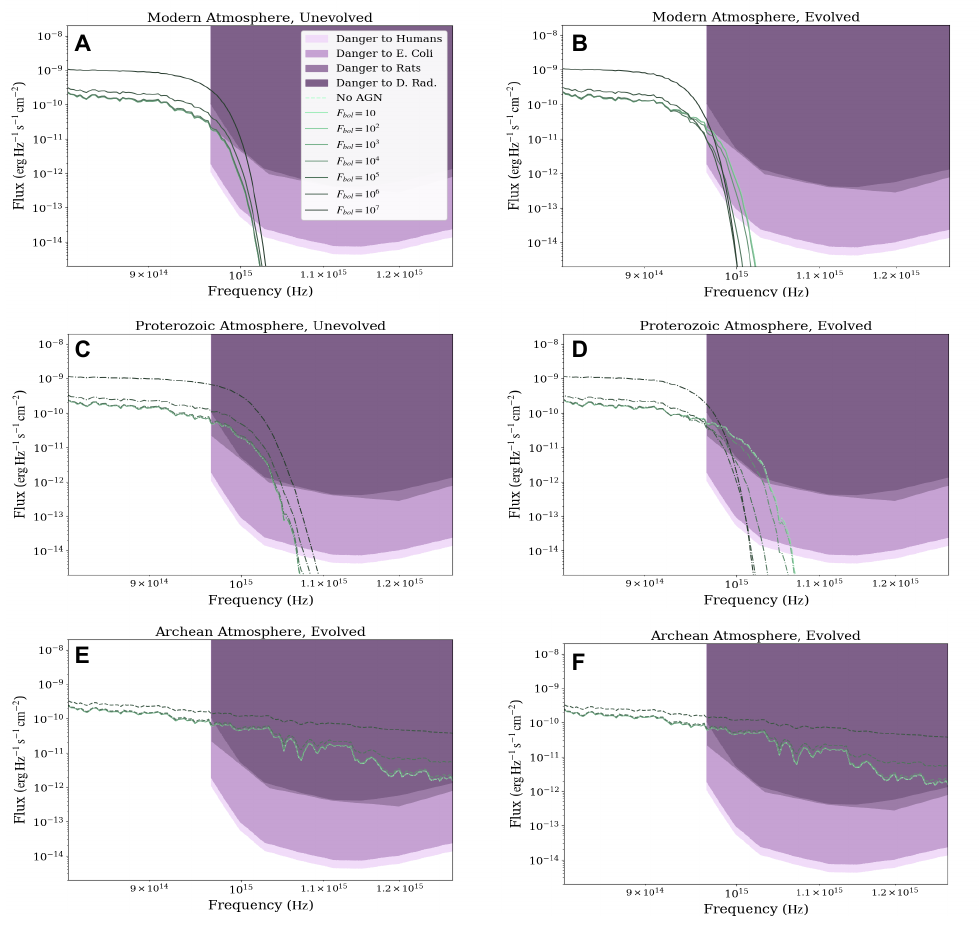}
    \caption{Surface flux in $\rm erg\,s^{-1}\,cm^{-2}\,Hz^{-1}$ versus frequency in $\rm Hz$ plots, with danger spectrum for species shaded in purple. Panels A, C, and E: Initial atmospheric conditions. Panels B, D, and F: Evolved atmospheric conditions, simulated in \texttt{PALEO} (same as lower three panels of Figure \ref{fig:all_flux}). Panels A and B: Modern atmosphere. We see that substantial protection from UV radiation is developed after atmospheric evolution, with higher AGN fluxes generating greater UV protection than the Sun-only case. Panels C and D: Proterozoic atmosphere. The improvement in atmospheric protection from UV radiation is larger in magnitude here, but ultimately less protection is provided than in the same simulations under the Modern atmosphere. Panels E and F: Archean atmosphere. Very little atmospheric protection against UV radiation develops.}
    \label{fig:all_sidebyside}
\end{figure*}

To determine the reduction of danger that species face as a result of their planetary atmospheres evolving, we plot the \texttt{PALEO} surface flux output over our limits of danger to species established previously. These are shown side-by-side with the same plot for the initial/unevolved atmospheres in Figure \ref{fig:all_sidebyside}. 

\begin{figure}
    \centering
    \includegraphics[scale=0.38]{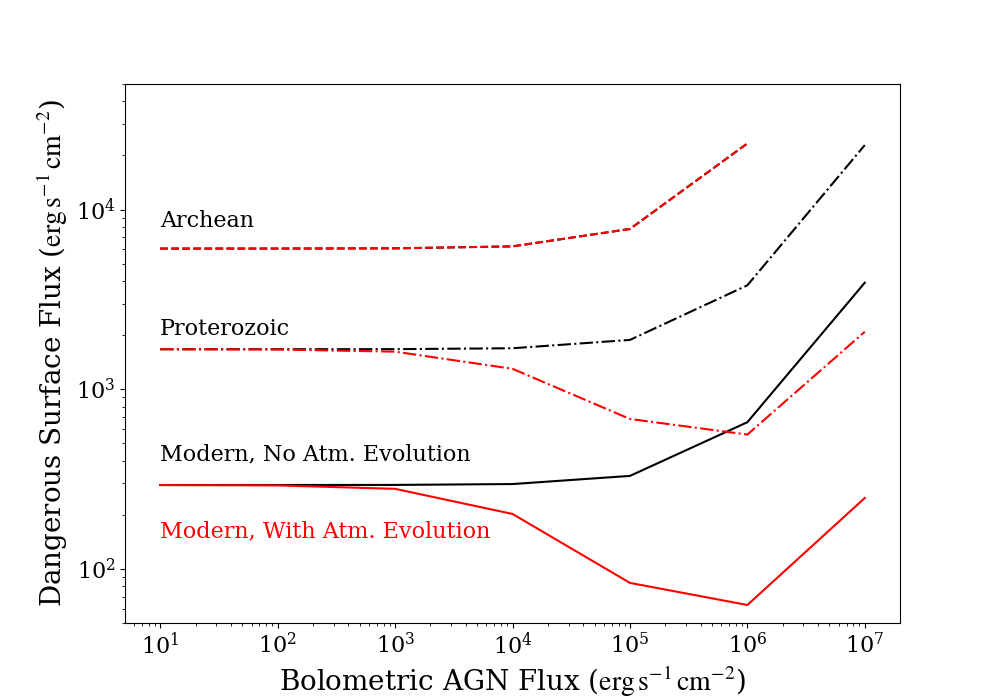}
    \caption{Integrated UV flux posing a danger to living species on the planetary surface, versus bolometric TOA AGN flux. Both axes are in units of $\rm erg \, s^{-1} \, cm^{-2}$. Here, dangerous surface flux is considered to be when any part of the shaded purple regions shown in Figure \ref{fig:all_sidebyside} is beneath the total AGN plus solar flux curve. These integrated fluxes correspond to the area between the surface AGN plus solar flux curves and the bottom-most purple flux curve which represents danger to human skin cells. The very low AGN flux cases are equivalent to the models including Sun's radiation with no AGN. We can see clearly that when we account for evolution of the atmosphere in the Modern and Proterozoic cases, the dangerous UV flux on the surface decreases with increasing AGN flux up until $10^6 \rm\, erg \, s^{-1} \, cm^{-2}$. The dangerous surface flux then begins to rise again at $10^7 \rm\, erg \, s^{-1} \, cm^{-2}$, but is still comparable to or less than the initial scenario where the AGN flux was negligible. For the Archean atmospheres, the scenarios with and without atmospheric evolution are virtually identical.}
    \label{fig:money}
\end{figure}

With the presence of a stellar spectrum (the Sun), we see that all species under a given atmosphere do experience some level of danger from stellar radiation--- even with no AGN present--- as quantified by the purple zones of `species danger' which are plotted over each panel of Figure \ref{fig:all_sidebyside}, and were motivated in Section \ref{sec:UVspecies}. Nevertheless, there are variations in how much area of the purple-shaded danger zone is included under the different flux curves, and therefore the integrated dangerous flux received at the surface. Figure \ref{fig:all_sidebyside} shows differential flux $F_\nu$ versus frequency $\nu$, so an area shown on a plot can be expressed as an integrated flux. We can qualitatively compare the total amount of dangerous flux received in each simulation to make statements about relative protection received from the atmosphere. The amount of dangerous flux received versus overall AGN flux is also quantitatively represented in Figure \ref{fig:money}.

In the modern atmosphere case, Figure \ref{fig:all_sidebyside}A and B, we see a smaller area of the purple shaded regions are included under the flux curves with an evolved atmosphere versus the initial atmosphere, indicating a reduction in danger to species. Notably, the reduction is more pronounced with the higher flux cases. This is consistent with a scenario where larger amounts of UV flux generate higher ozone mixing ratios and higher levels of protection on the surface at lower UV wavelengths. In Figure \ref{fig:money}, we see that even in the most extreme AGN flux case we test, there is still less dangerous flux on the planetary surface versus a scenario where solar flux dominates. 

In the Proterozoic atmosphere case, Figure \ref{fig:all_sidebyside}C and D, we see that under the evolved atmosphere the lowest flux curves cover a substantially larger area on the shaded purple zone where species are in danger versus the higher flux cases. This is similar to what was observed in the modern atmosphere, however the transition is more dramatic in this case, with a lower amount of starting oxygen (surface mixing ratio of $10^{-3}$) generating less initial protection. However, the level of dangerous flux on the surface is still much more than the same AGN flux scenarios for the Modern atmosphere, as we see in Figure \ref{fig:money}.

Both the Proterozoic and Modern atmospheres exhibit slightly increased protection overall in the higher flux cases compared not only to the lower AGN flux cases, but also compared to the No AGN case. This indicates that the production of additional ozone triggered by the AGN radiation actually results in a net increase in surface UV protection for species, as compared to the same atmosphere with only the Sun's radiation. This raises the interesting question of how the presence of an AGN may actually foster improved habitability in the long-term, after potential short-term damages to species from UV radiation.    

Lastly, in the Archean atmosphere case shown in Figure \ref{fig:all_sidebyside}E and F, we see virtually no change between the initial and evolved atmospheres. This is consistent with what we saw in Section \ref{sec:archatm}. During the actual Archean period, life on Earth consisted of monocellular organisms, which were more resistant to radiation than many modern species \citep{Lep2020}. Nevertheless, our results shows that even \textit{D. Radiodurans}, one of the most radiation-resistant monocellular organisms known today, is still in danger on the surface of a planet facing these conditions--- so even very hardy, early life developing on a planet's surface would still be at risk. Oceans may provide further protection from UV radiation that allows life to persist underwater, as demonstrated in \citet{Estrela2020} for exoplanets experiencing high UV radiation from M dwarf flares, but that is beyond the scope of this work.  

\subsubsection{Stellar Population Impacts in Galaxies} \label{sec:agnresults}
Previous works on the subject of AGN UV radiation and galactic habitability \citep{BBB2019, Pac2020} have considered only the radiation from the AGN and not the additional radiation of a host star, and given distances from the galactic center within which a hypothetical planet would become dangerous to life due to high UV radiation. In order to make this work more naturally comparable to these previous studies, we now analyze the results found in Section \ref{sec:finalresults} using the AGN flux only.

\begin{figure*}
    \centering
    \includegraphics[scale=1]{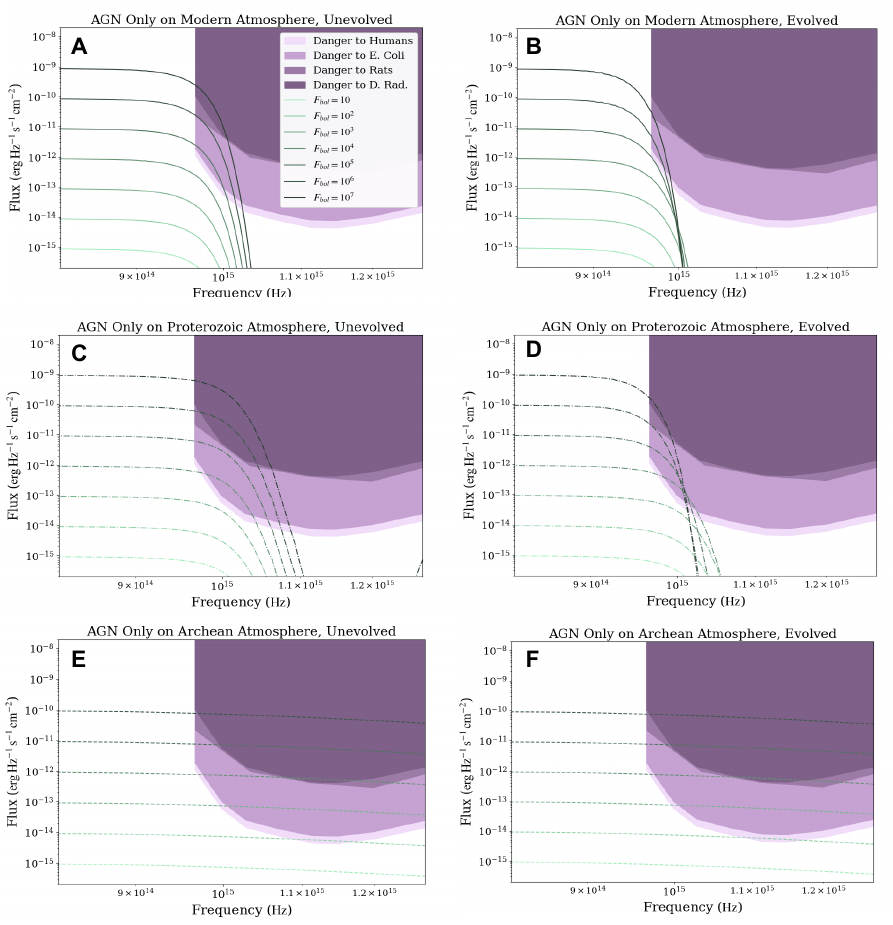}
    \caption{Surface flux in $\rm erg\,s^{-1}\,cm^{-2}\,Hz^{-1}$ versus frequency in $\rm Hz$ plots for the AGN flux only, with danger spectrum for species shaded in purple. Identical to Figure \ref{fig:all_sidebyside} with solar flux subtracted out. Panels A and B: Modern atmosphere. Panels C and D: Proterozoic atmosphere. Panels E and F: Archean atmosphere. Panels A, C, and E: Initial atmospheric conditions. Panels B, D, and F: Evolved atmospheric conditions, simulated in \texttt{PALEO}. Note the different scale of the y-axis to Figure \ref{fig:all_sidebyside}. The same trends as in \ref{fig:all_sidebyside} are observed for all three atmospheres.}
    \label{fig:all_agnonly}
\end{figure*}

The surface flux \textit{from only the AGN} before and after chemical evolution of the atmosphere is shown in Figure \ref{fig:all_agnonly}. This is identical to Figure \ref{fig:all_sidebyside} but with the solar flux subtracted out. The point at which a species is in danger is now defined as the radial distance from the galactic center at which the surface differential flux ($F_\nu$) from the AGN exceeds the lower limit of danger for the species at any frequency. We determine the bolometric flux at which this occurs by linearly interpolating between the surface flux SEDs shown in Figure \ref{fig:all_agnonly} \citep{2020SciPy-NMeth}. These flux limits representing danger to species in a given galaxy are shown in Table \ref{tab:fluxes321}. The distances from the galactic center within which species face danger are shown in Table \ref{tab:m87} for M87, Table \ref{tab:mw}, and Table \ref{tab:ngc1277}, as well as in corresponding tables in Appendix Section \ref{sec:rednuggets} for the other red nugget relic galaxies. Also listed in these tables are the percentages of the stellar population in a given galaxy which would experience equal to or greater than the dangerous level of AGN flux. 

We are not able to find exact distance or population percentage values for \textit{D. Radiodurans} under a modern atmosphere, as the entire shaded danger region to \textit{D. Radiodurans} (Figure \ref{fig:all_agnonly}B) sits above the highest AGN flux simulation that converged under the Modern atmosphere, so no interpolation could be performed. However the flux needed to cause danger to \textit{D. Radiodurans} is greater than that needed to cause danger to rats across all of the UV wavelengths we interrogate, so we list the values derived for rats as upper or lower limits.

\begin{table}[h]
    \centering
    \begin{tabular}{cc||c|c|c|c|} 
    \hline
    \textbf{\footnotesize Atmosphere} & \textbf{\footnotesize Time} & \footnotesize Human & \footnotesize \textit{E. Coli}& \footnotesize Rat & \footnotesize \textit{D. Rad.}\\ \Xhline{2\arrayrulewidth}
    
    \multirow{2}{4em}{Modern} & $t=0$ & $10^{4.15}$ &  $10^{4.39}$ & $10^{5.68}$ & $10^{6.15}$\\ \cline{2-6}
    
    &$t>36$ d & $10^{5.05}$ & $10^{5.40}$ & $10^{6.82}$ & $>10^{6.82*}$\\  \Xhline{2\arrayrulewidth}

    \multirow{2}{4em}{Proterozoic} & $t=0$ & $10^{3.03}$ & $10^{3.26}$ & $10^{4.94}$ & $10^{5.01}$ \\ \cline{2-6}

    & $t > 36$ d & $10^{3.34}$ & $10^{3.62}$ & $10^{6.00}$ & $10^{6.71}$\\ \Xhline{2\arrayrulewidth}

    \multirow{2}{4em}{Archean} & $t=0$ & $10^{1.80}$ & $10^{2.03}$ & $10^{3.69}$ & $10^{3.76}$\\ \cline{2-6}

    & $t>36$ d & $10^{1.91}$ & $10^{2.14}$ & $10^{3.81}$ & $10^{3.90}$\\ \Xhline{2\arrayrulewidth}

    \end{tabular}
    \caption{\centering Bolometric AGN fluxes (in $\rm erg \, s^{-1}\, cm^{-2}$) necessary for a dangerous level of flux to various species be received on a planetary surface. Results are shown for various atmospheric compositions, before and after the atmospheric response to UV radiation from the AGN occurs. This table contextualizes the values presented in Tables \ref{tab:m87}, \ref{tab:mw}, \ref{tab:ngc1277}, as well as the analogous tables for the red nugget relic galaxies shown in Appendix Section \ref{sec:rednuggets}. Asterisked values are lower limits.}
    \label{tab:fluxes321}
\end{table}

In making broad conclusions about the impact of AGN on life across a galaxy, we are assuming Sunlike stars with Earthlike planets (with a corresponding likelihood of hosting life) are evenly distributed throughout each galaxy. We have included all species for all atmosphere types for consistency, but it should be noted that in Earth's history, only monocellular species existed during the Archean period \citep{Lep2020}, and species as complex as rats or humans only arose during the Modern period.

While significant radial distances from the galactic center are in danger during Eddington-limited accretion in M87 before and after atmospheric evolution, particularly for the Proterozoic and Archean atmospheres, the corresponding stellar populations are still very small (see Table \ref{tab:m87}). Even the maximum possible radius of danger listed--- which is near $10 \rm\, kpc$, for humans under the Archean atmosphere before atmospheric evolution--- accounts for less than half of the total stellar population. In spite of this large elliptical galaxy having a more centrally peaked stellar population as compared to spiral galaxies, its enormous size still means less than half of its total stars are contained in the inner regions ($< 10 \rm\, kpc$ from the center), so the majority of potential planetary systems in the galaxy are always safe from this radiation. 

The decrease in danger to species on planetary surfaces is substantial under the Modern and Proterozoic atmospheres. For the Modern atmosphere, the fraction of the stellar population which would have its planets affected by dangerous radiation levels before atmospheric evolution is up to $10$ times greater than after atmospheric evolution. For Proterozoic, the population affected before atmospheric evolution is up to $20$ times than after. The changes in the Archean atmosphere are more subtle, but there is still a small decrease in the percentage of potential planets experiencing dangerous levels of UV flux on their surfaces. 

In the Milky Way bulge, we observe that the radii of danger during Eddington-limited accretion are mostly insignificant compared to the extent of the bulge itself \citep[around $3 \rm\, kpc$, per][]{NewMWSofue2009}, but this still accounts for percentages of the stellar population in the bulge comparable the same scenario in M87 (see Table \ref{tab:mw}). If so little of the bulge is affected, assuming zero attenuation by the ISM, then the vast majority of the stellar population in the disk, which contains the majority of its stars within a radius of roughly $10 \rm\, kpc$ but has a full extent out to $20 \rm\, kpc$ \citep{NewMWSofue2009}, as well as a substantial ISM, will be unaffected. We also see a similar result to M87, where there are substantial decreases in the percent of planets receiving dangerous surface flux after atmospheric evolution.

\begin{table*}[t]
    \centering
    \begin{tabular}{ccc||cc|cc|cc|cc|cc|} 
    \hline
    \multicolumn{3}{c}{\textbf{\footnotesize Species}} & \multicolumn{2}{c}{\footnotesize Human} & \multicolumn{2}{c}{\footnotesize \textit{E. Coli}}& \multicolumn{2}{c}{\footnotesize Rat} & \multicolumn{2}{c}{\footnotesize \textit{D. Radiodurans}}\\ \hline
    \textbf{\footnotesize Atmosphere} & \textbf{\footnotesize Time} &\textbf{\footnotesize Edd. Ratio} & \footnotesize Distance (pc) & \footnotesize{Population} & \footnotesize Dist. &  \footnotesize{Pop.} & \footnotesize Dist. &  \footnotesize{Pop.} & \footnotesize Dist. &  \footnotesize{Pop.} \\ \Xhline{2\arrayrulewidth}
    
    \multirow{6}{4em}{Modern} & \multirow{3}{4em}{$t=0$} & $1$ & $662$ & $2.0\%$& $501$ & $1.1\%$ & $113$ & $0.05\%$& $66$ & $0.02 \% $\\
    &&$0.1$ & $208$ & $0.2\%$& $157$ & $0.1\%$ & $35$ & $0.01\%$ & $21$ & $<0.01\% $ \\ 
    && $0.01$ & $65$ & $0.02\%$  & $49$ & $0.01\%$ & $11$ & $< 0.01\%$ & $7$ &$<0.01\% $ \\  \cline{2-11}
    
    & \multirow{3}{4em}{$t>36$ d}& $1$ & $234$ & $0.2\%$& $157$ & $0.1\%$ & $30$ & $<0.01\%$& $<30^*$ & $<0.01 \%$\\ 
    &&$0.1$ & $74$ & $0.02\%$& $50$ & $0.01\%$ & $10$ & $<0.01\%$ & $<10^*$ & $<0.01\%$ \\ 
    && $0.01$ & $23$ & $<0.01\%$  & $16$ & $<0.01\%$ & $3$ & $< 0.01\%$ & $<3^*$ &$<0.01\% $ \\ \Xhline{2\arrayrulewidth}

    \multirow{6}{4em}{Proterozoic} & \multirow{3}{4em}{$t=0$} & $1$ & $2397$ & $12.8\%$ & $1836$ & $9.5\%$ & $202$ & $0.2\%$ &$247$ & $0.2\%$\\
    &&$0.1$ & $765$ & $2.6\%$ &$584$ & $1.5\%$ & $63$ & $0.02\%$ & $78$ & $0.03\%$  \\ 
    && $0.01$ & $243$ & $0.2\%$ & $185$ & $0.1\%$ & $20$ & $<0.01\%$  & $25$ & $<0.01\%$ \\ \cline{2-11}

    &\multirow{3}{4em}{$t > 36$ d} & $1$ & $1673$ & $8.5\%$ & $1218$ & $5.6\%$ & $78$ & $0.03\%$ &$35$ & $<0.01\%$\\
    &&$0.1$ & $529$ & $1.3\%$ &$385$ & $0.6\%$ & $25$ & $<0.01\%$ & $11$ & $<0.01\%$  \\ 
    && $0.01$ & $167$ & $0.1\%$ & $122$ & $0.06\%$ & $8$ & $<0.01\%$  & $3$ & $<0.01\%$ \\ \Xhline{2\arrayrulewidth}

    \multirow{6}{4em}{Archean} & \multirow{3}{4em}{$t=0$} & $1$ &$9927$ & $41.2\%$ & $7560$ & $34.1\%$ & $1126$ & $4.9\%$ & $1028$ & $4.3\%$\\
    &&$0.1$ & $3155$ & $16.8\%$ & $2420$ & $13.0\%$ & $355$ & $0.5\%$ & $327$ & $0.5\%$\\
    && $0.01$ & $1009$ & $4.2\%$ & $772$ & $2.6\%$ & $112$ & $0.05\%$ & $104$ & $0.05\%$ \\ \cline{2-11}

    &\multirow{3}{4em}{$t>36$ d} & $1$ &$8764$ & $37.9\%$ & $6674$ & $31.1\%$ & $982$ & $4.0\%$ & $887$ & $3.4\%$\\
    &&$0.1$ & $2772$ & $14.8\%$ & $2110$ & $11.2\%$ & $311$ & $0.4\%$ & $281$ & $0.3\%$\\
    && $0.01$ & $876$ & $3.3\%$ & $667$ & $2.0\%$ & $98$ & $0.04\%$ & $89$ & $0.03\%$ \\ \Xhline{2\arrayrulewidth}

    \end{tabular}
    \caption{\centering \textbf{M87}: \footnotesize Physical situations corresponding to the bolometric AGN fluxes given in Table \ref{tab:fluxes321} in the M87 galaxy. Greatest radial distances (in $\rm pc$) from the center of M87 which experience the level of flux dangerous to humans, \textit{E. Coli}, rats, and \textit{D. Radiodurans}, at various Eddington ratios of M87* and assuming different planetary atmospheres. Results are shown for before and after the atmospheric response to UV radiation from the AGN occurs. This table also includes the percentage of the stellar population of M87 receiving equal or greater to a dangerous level of flux. Asterisked values are upper limits.}
    \label{tab:m87}
\end{table*}

\begin{table*}[t]
    \centering
    \begin{tabular}{ccc||cc|cc|cc|cc|cc|} 
    \hline
    \multicolumn{3}{c}{\textbf{\footnotesize Species}} & \multicolumn{2}{c}{\footnotesize Human} & \multicolumn{2}{c}{\footnotesize \textit{E. Coli}}& \multicolumn{2}{c}{\footnotesize Rat} & \multicolumn{2}{c}{\footnotesize \textit{D. Radiodurans}}\\ \hline
    \textbf{\footnotesize Atmosphere} & \textbf{\footnotesize Time} &\textbf{\footnotesize Edd. Ratio} & \footnotesize Distance (pc) & \footnotesize{Population} & \footnotesize Dist. &  \footnotesize{Pop.} & \footnotesize Dist. &  \footnotesize{Pop.} & \footnotesize Dist. &  \footnotesize{Pop.} \\ \Xhline{2\arrayrulewidth}
    
    \multirow{6}{4em}{Modern} & \multirow{3}{4em}{$t=0$} & $1$ & $18$ & $1.5\%$ & $13$ & $1.1\%$ & $3$ & $0.2\%$ & $2$ & $0.1\%$ \\
     &&$0.1$ & $6$ & $0.5\%$ & $4$ & $0.3\%$ & $0.9$ & $0.1\%$& $0.6$ & $0.05\%$ \\
    && $0.01$ & $2$ & $0.1\%$ & $1$ & $0.1\%$ & $0.3$ & $0.02\%$ & $0.2$ & $0.01\%$ \\ \cline{2-11}
    
    & \multirow{3}{4em}{$t>36$ d}& $1$ & $6$ & $0.5\%$ & $4$ & $0.3\%$ & $0.8$ & $0.07\%$ & $<0.8^*$ & $<0.07\%^*$\\
     &&$0.1$ & $2$ & $0.2\%$ & $1$ & $0.1\%$ & $0.2$ & $0.02\%$& $<0.2^*$ & $<0.02\%^*$ \\
    && $0.01$ & $0.6$ & $0.05\%$ & $0.4$ & $0.03\%$ & $0.08$ & $<0.01\%$ & $<0.08^*$ & $<0.01\%$ \\ \Xhline{2\arrayrulewidth}

    \multirow{6}{4em}{Proterozoic} & \multirow{3}{4em}{$t=0$} & $1$ & $63$ & $5.3\%$ & $48$ & $4.0\%$ & $5$ & $0.4\%$ & $6$ & $0.5\%$\\
    &&$0.1$ & $20$ & $1.6\%$ & $15$ & $1.3\%$ & $2$ & $0.1\%$  & $2$ & $0.2\%$ \\
    && $0.01$ & $6$ & $0.5\%$ & $5$ & $0.4\%$
    & $0.5$ & $0.04\%$  & $0.6$ & $0.05\%$ \\ \cline{2-11}

    &\multirow{3}{4em}{$t > 36$ d} & $1$ & $44$ & $3.7\%$ & $32$ & $2.7\%$ & $2$ & $0.2\%$ & $0.9$ & $0.08\%$\\
    &&$0.1$ & $14$ & $1.2\%$ & $10$ & $0.8\%$ & $0.6$ & $0.05\%$  & $0.3$ & $0.02\%$ \\
    && $0.01$ & $4$ & $0.4\%$ & $3$ & $0.3\%$ & $0.2$ & $0.02\%$  & $0.09$ & $<0.01\%$ \\ \Xhline{2\arrayrulewidth}

    \multirow{6}{4em}{Archean} & \multirow{3}{4em}{$t=0$} & $1$ &$265$ & $23.3\%$ & $203$ & $17.4\%$ & $29$ & $2.4\%$ & $27$ & $2.2\%$ \\
    &&$0.1$ & $83$ & $7.0\%$ & $64$ & $5.3\%$ & $9$ & $0.8\%$ & $8$ & $0.7\%$\\
    && $0.01$ & $26$ & $2.2\%$ & $20$ & $1.7\%$ & $3$ & $0.2\%$ & $3$ & $0.2\%$\\ \cline{2-11}

    &\multirow{3}{4em}{$t>36$ d} & $1$ &$231$ & $20.0\%$ & $176$ & $14.8\%$ & $26$ & $2.2\%$ & $23$ & $2.0\%$\\
    &&$0.1$ & $73$ & $6.1\%$ & $56$ & $4.7\%$ & $8$ & $0.7\%$ & $7$ & $0.6\%$\\
    && $0.01$ & $23$ & $1.9\%$ & $18$ & $1.5\%$ & $3$ & $0.2\%$ & $2$ & $0.2\%$\\ \Xhline{2\arrayrulewidth}

    \end{tabular}
    \caption{\centering \textbf{Milky Way Bulge}: \footnotesize Physical situations corresponding to the bolometric AGN fluxes given in Table \ref{tab:fluxes321} in the Milky Way bulge. Greatest radial distances (in $\rm pc$) from the center of the Milky Way which experience the level of flux dangerous to humans, \textit{E. Coli}, rats, and \textit{D. Radiodurans}, at various Eddington ratios of Sag A* and assuming different planetary atmospheres. Results are shown for before and after the atmospheric response to UV radiation from the AGN occurs. This table also includes the percentage of the stellar population of the Milky Way bulge receiving equal or greater to a dangerous level of flux. Asterisked values are upper limits.}
    \label{tab:mw}
\end{table*}

The red nugget relic galaxies show an even more centrally peaked stellar population than the large elliptical M87. While the physical regions of the galaxy experiencing potentially dangerous radiation in NGC 1277 during Eddington-limited accretion are similar in size to those in M87, since both have SMBH masses on the order of $10^9 M_\odot$, we see that the corresponding percentage of stars affected is much higher in NGC 1277 than in M87. In particular, for several species and Eddington ratios under an Archean atmosphere we see that the majority of stars in the galaxy are affected, and there is only a slight decrease in danger to species after atmospheric evolution, as is consistent with our results for the Milky Way and M87. This suggests that species in earlier eras of the Universe, whose galaxies may have looked more like these relics \citep[e.g.][]{VanD2010, Barro2013, Vand2014}, could have been in significantly more danger from UV radiation from AGN than species in galaxies we typically see today. Depending on the shielding properties of individual planetary atmospheres, even worlds at the furthest edge of the galaxy could still be affected. This may have impacted the formation or continuation of existing life in the inner regions of red nugget galaxies for significant parts of their history.

\begin{table*}[t]
    \centering
    \begin{tabular}{ccc||cc|cc|cc|cc|cc|} 
    \hline
    \multicolumn{3}{c}{\textbf{\footnotesize Species}} & \multicolumn{2}{c}{\footnotesize Human} & \multicolumn{2}{c}{\footnotesize \textit{E. Coli}}& \multicolumn{2}{c}{\footnotesize Rat} & \multicolumn{2}{c}{\footnotesize \textit{D. Radiodurans}}\\ \hline
    \textbf{\footnotesize Atmosphere} & \textbf{\footnotesize Time} &\textbf{\footnotesize Edd. Ratio} & \footnotesize Distance (pc) & \footnotesize{Population} & \footnotesize Dist. &  \footnotesize{Pop.} & \footnotesize Dist. &  \footnotesize{Pop.} & \footnotesize Dist. &  \footnotesize{Pop.} \\ \Xhline{2\arrayrulewidth}
    
    \multirow{6}{4em}{Modern} & \multirow{3}{4em}{$t=0$} & $1$ & $609$ & $22.0\%$ & $463$ & $16.4\%$ & $104$ & $2.7\%$& $61$ & $1.3\%$\\ 
     &&$0.1$ & $192$ & $5.8\%$ & $146$ & $4.1\%$ & $33$ & $0.5\%$ & $19$ & $0.2\%$ \\ 
    && $0.01$ & $61$ & $1.3\%$ & $46$ & $0.9\%$ & $10$ & $<0.04\%$ & $6$ & $<0.04\%$  \\ \cline{2-11}
    
    & \multirow{3}{4em}{$t>36$ d}& $1$ & $213$ & $6.6\%$ & $143$ & $4.0\%$ & $28$ & $0.4\%$& $<28^*$ & $<0.4\%^*$\\ 
     &&$0.1$ & $67$ & $1.5\%$ & $45$ & $0.9\%$ & $9$ & $<0.04\%$ & $<9^*$ & $<0.04\%$ \\ 
    && $0.01$ & $21$ & $0.2\%$ & $14$ & $0.05\%$ & $3$ & $<0.04\%$ & $<3^*$ & $<0.04\%$  \\ \Xhline{2\arrayrulewidth}

    \multirow{6}{4em}{Proterozoic} & \multirow{3}{4em}{$t=0$} & $1$ & $2193$ & $58.8\%$ &$1658$ & $50.0\%$& $185$ & $5.5\%$ & $220$ & $6.9\%$\\
    &&$0.1$ & $688$ & $25.0\%$ &$519$ & $18.6\%$& $59$ & $1.3\%$ & $70$ & $1.6\%$  \\ 
    && $0.01$ & $215$ & $6.7\%$ &$165$ & $4.7\%$ & $18$ & $0.1\%$ & $22$ & $0.2\%$ \\ \cline{2-11}

    &\multirow{3}{4em}{$t > 36$ d} & $1$ & $1525$ & $47.6\%$ & $1110$ & $38.3\%$& $71$ & $1.6\%$ & $32$ & $0.5\%$\\
    &&$0.1$ & $482$ & $17.2\%$ &$351$ & $12.1\%$& $23$ & $0.2\%$ & $10$ & $<0.04\%$  \\ 
    && $0.01$ & $152$ & $4.3\%$ &$111$ & $2.9\%$ & $7$ & $<0.04\%$ & $3$ & $<0.04\%$ \\ \Xhline{2\arrayrulewidth}

    \multirow{6}{4em}{Archean} & \multirow{3}{4em}{$t=0$} & $1$ & $9264$ & $99.9\%$ &$7018$ & $95.2\%$& $1015$ & $35.8\%$ & $924$ & $33.2\%$\\
     &&$0.1$ & $2913$ & $68.7\%$ &$2204$ & $59.0\%$& $323$ & $11.0\%$ & $289$ & $9.7\%$ \\
    && $0.01$ & $914$ & $32.9\%$ &$691$ & $25.1\%$& $102$ & $2.6\%$ & $93$ & $2.3\%$\\\cline{2-11}

    &\multirow{3}{4em}{$t>36$ d} & $1$ & $7987$ & $97.9\%$ &$6082$ & $91.6\%$& $895$ & $32.3\%$ & $809$ & $29.4\%$\\
     &&$0.1$ & $2526$ & $63.7\%$ &$1923$ & $54.5\%$& $283$ & $9.4\%$ & $256$ & $8.3\%$ \\
    && $0.01$ & $799$ & $29.1\%$ &$608$ & $22.0\%$& $89$ & $2.2\%$ & $81$ & $1.9\%$\\ \Xhline{2\arrayrulewidth}

    \end{tabular}
    \caption{\centering \textbf{NGC 1277}: \footnotesize Physical situations corresponding to the bolometric AGN fluxes given in Table \ref{tab:fluxes321} in the NGC 1277 galaxy. Greatest radial distances (in $\rm pc$) from the center of NGC 1277 which experience the level of flux dangerous to humans, \textit{E. Coli}, rats, and \textit{D. Radiodurans}, at various Eddington ratios of NGC 1277* and assuming different planetary atmospheres. Results are shown for before and after the atmospheric response to UV radiation from the AGN occurs. This table also includes the percentage of the stellar population of NGC 1277 receiving equal or greater to a dangerous level of flux. Asterisked values are upper limits.}
    \label{tab:ngc1277}
\end{table*}

\subsection{Discussion}

Overall, these results suggest that in atmospheres with initial oxygen surface mixing ratio $\geq 10^{-3} \rm\, mol/mol$, AGN radiation would generate an increase in protection against UV radiation on the planetary surface due to the development of a thicker ozone layer. This evolution is more dramatic when the AGN flux received is higher--- i.e. for worlds nearer to the galactic center, or when the AGN is radiating at a higher Eddington ratio. In all cases, the net level of dangerous UV radiation on the surface is less than or comparable to the No AGN/baseline conditions. However, if the level of oxygen in the initial atmosphere is not adequate, as with the Archean, there is virtually no ozone evolution as a result of AGN flux, and all of the excess flux from the AGN goes directly to the planetary surface. Our model is also cloud-free, and the clouded areas in simulations such as those of \citet{Ridgway2023} experienced a substantial increase in UV protection on the planetary surface, so it is possible that with the presence of clouds the protection from UV radiation could be heightened even further.

Planets in the innermost regions of the galaxy with atmospheres like the Modern and Proterozoic types would gain the most protection. This means we can plausibly envision a scenario where we see a ``ring'' of danger about the galactic center, i.e. there may be some ``goldilocks zone'' near the AGN where AGN radiation has net neutral or positive effects for habitability via production of ozone and increased protection on the surface from UV radiation, and a zone of danger outside of that. This is an interesting question to explore in future work.

It is also possible that atmospheric evaporation caused by AGN (via XUV radiation or charged particle winds) could counteract the protective effects of increased ozone in a planet's atmosphere. \citet{Amb2022} show that winds from Sagittarius A* in an AGN phase have the potential to cause the depletion of up to $99\%$ of an Earth-like planet's atmospheric ozone at a distance of $\sim 100 \rm\, pc$, and this effect could be more substantial in M87 or red nugget relic galaxies, which have larger central SMBHs and denser stellar populations near their galactic centers. Due to its particular effect on ozone, this process may temper the strongest protective effects that we saw near the inner regions of these galaxies. XUV radiation can cause a similar effect, and \citet{Wis2019} found that Sagittarius A* in an AGN phase may evaporate up to several Earth-atmosphere-masses of atmospheric gas from MW bulge worlds. This may likewise potentially temper the protective effects of ozone production. However, it has also been suggested that atmospheric evaporation may make some non-terrestrial planets more habitable, by converting `mini-Neptunes' with hydrogen- and helium-based atmospheres into super-Earths \citep{Luger2015,Chen2018}.

Our work exhibits a ``Gaian bottleneck'' phenomenon, where life is very vulnerable to potential extinction events before it has substantially altered its planetary atmosphere towards more suitable conditions for itself, and after that point is invulnerable to similar potential extinction events \citep[e.g.][]{Love1974, Nich2018,Alca2020, Nich2023}. In our Modern and Proterozoic models (which correspond to epochs on Earth with more complex life than during the Archean), not only is the atmosphere able to recover its original surface flux conditions after being perturbed by radiation from an AGN, but the atmosphere now tends towards becoming \textit{even more} resistant to future UV irradiating events than it was originally, as demonstrated in Figure \ref{fig:money}. On the other hand, on many Earth-like worlds with an Archean atmosphere, even one of the most radiation-resistant organisms that exists today, \textit{D. Radiodurans}, would experience a harmful level of radiation. When UV radiation from an AGN strikes at a more vulnerable point in a planet's history (where the atmosphere is not yet oxygenated), the progression of life may be halted entirely, preventing those planets from becoming hospitable to life in the future.

\section{Conclusion}
We have studied the impact of UV radiation from an AGN on the surface habitability of planets and on their atmospheric composition. We find that for galaxies similar to M87 and the Milky Way, only the innermost regions of the galaxy are dramatically affected by high UV radiation from an AGN, and even then the affected volume typically contains a low percentage of the galaxy's overall stellar population. This is the case even in massive ellipticals such as M87, which have been proposed as potential candidates to be sterilized by AGN radiation in previous works due to their more centrally peaked stellar population density \citep{whit2020}. In spiral galaxies such as the Milky Way, the level of danger is even lower, seeing as only the bulge population is affected, even when we completely disregard attenuation by the ISM. In the even \textit{more} centrally-peaked populations of the red nugget relic galaxies, AGN radiation can affect a larger percentage of the overall potential planetary systems--- potentially even the majority of planetary systems in the galaxy, depending on the atmospheric oxygen levels of individual planets. This indicates that galaxies being sterilized by AGN radiation could have been a more important phenomenon in the early Universe than it is in the typical present-day galaxies we considered (M87 and the Milky Way).   

We find that the level of danger on individual planetary surfaces is highly dependent on the atmospheric conditions, particularly the level of oxygen. For planets with substantial atmospheric oxygen (mixing ratio greater than roughly $10^{-3} \rm\, mol/mol$), AGN radiation could have a neutral or positive effect on habitability in the galaxy rather than discourage it, particularly in the inner regions of the galaxy. However on planets with lower atmospheric oxygen, no ozone layer can develop and the level of danger remains fairly consistent over time. A runaway greenhouse effect is predicted to occur in the two highest flux cases we model, which could potentially lead to a net detriment to habitability on planets with oxygenated atmospheres receiving high levels of flux from an AGN.

Working out the nuances of where this increased-habitability effect eventually gives way to a runaway greenhouse, and what percentages of a stellar population may benefit from or be harmed by these effects, remains another interesting question to explore in the future. One exciting possibility for future work is to include UV radiation from an AGN in a 3D climate simulation, which would be equipped to model spatial and climate effects on exoplanets that are not possible in a 1D simulation, and--- importantly for our results here--- to model a runaway greenhouse scenario \citep[e.g.][]{Mayne2014}. These simulations would facilitate an exploration of the potential secondary dangers to habitability that result from the AGN radiation.   

\section*{Acknowledgements}
We thank Jonathan H.\ Cohn for assistance with selecting the red nugget relic galaxy sample.
R.C.H.\ is grateful for support from NASA through Astrophysics Data Analysis Program grant 80NSSC23K0485. N.J.M.\ gratefully acknowledges funding from a Leverhulme Trust Research Project Grant [RPG-2020-82]. This work was partly supported by a Science and Technology Facilities Council Consolidated Grant [ST/R000395/1]. This work was also supported by a UKRI Future Leaders Fellowship [grant number MR/T040866/1].
\software{\texttt{PALEO} \citep{Daines2016, Jake2024}, \texttt{scipy} \citep{2020SciPy-NMeth}, \texttt{astropy} \citep{Astropy2018}, \texttt{numpy} \citep{Numpy}}

\bibliographystyle{aasjournal}
\bibliography{bib}

\appendix
\section{Additional Figures}

\subsection{Visual Representation of Fluxes Received at Various Distances from an AGN}

\begin{figure*}
    \centering
    \includegraphics[scale=0.35]{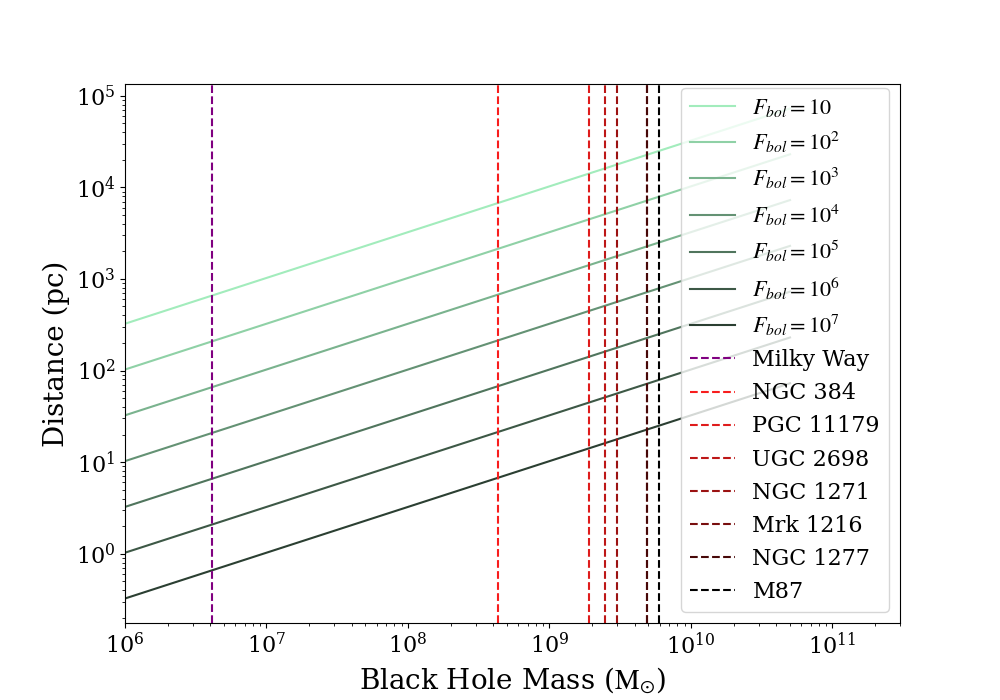}
    \caption{Visual representation of the $100\%$ Eddington ratio rows of Tables \ref{tab:scaling} and \ref{tab:scaling_app} (the latter shown in Appendix Section \ref{sec:rednuggets}). Solid green lines represent constant bolometric AGN fluxes used in our simulations ($10$, $10^2$, $10^3$, etc. through $10^7 \rm\, erg\, s^{-1} \, cm^{-2}$), with mass of a given black hole on the x-axis in $M_\odot$, and the radial distance from the BH at which that flux is received on the y-axis in $\rm pc$. Vertical dashed lines mark the mass of the central SMBH in our galaxies of interest. Note that Mrk 1216 and NGC 1277 have the same black hole mass. Distance values in the $100\%$ Eddington rows of Tables \ref{tab:scaling} and \ref{tab:scaling_app} correspond to the points where the vertical dashed lines representing BH mass in a given galaxy intersect the solid green lines representing a given flux.}
    \label{fig:straightlines}
\end{figure*}

A visual representation of the information presented in Table \ref{tab:scaling} is shown in Figure \ref{fig:straightlines}. This figure also contains a visual representation of the equivalent table (Table \ref{tab:scaling_app}, shown in Appendix Section \ref{sec:rednuggets}) for the red nugget relic galaxy sample. 

\subsection{Extrapolation of Species' UV Flux Tolerance} \label{sec:extrap}
\begin{figure*}
    \centering
    \includegraphics[scale=0.6]{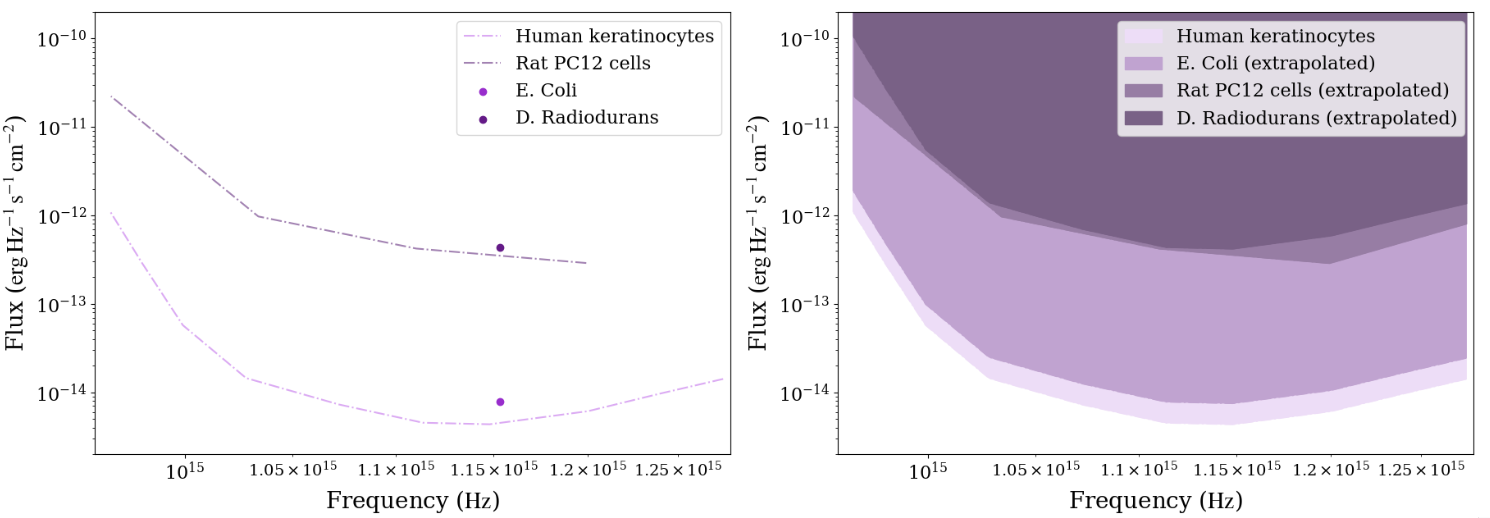}
    \caption{Limits of danger to various species, a differential flux $F_\nu$ in $\rm erg \, s^{-1} \,cm^{-2} \,Hz^{-1}$, as a function of frequency in $\rm Hz$. Human keratinocyte data comes from \citet{AokiHumans2011}, rat PC12 cell data from \citet{MasumaRats2013}, and \textit{E. Coli} and \textit{D. Radiodurans} from \citet{EColiDRad2010}. The left panel shows the original data, and the right panel shows our extrapolated version.}
    \label{fig:species_both}
\end{figure*}

The original data on species UV flux tolerance from \citet{AokiHumans2011}, \citet{MasumaRats2013}, and \citet{EColiDRad2010} is shown in the left panel of Figure \ref{fig:species_both}. The extrapolated limits of UV tolerance are shown in the right panel of Figure \ref{fig:species_both}.

\section{Red Nugget Relics} \label{sec:rednuggets}

\noindent The correspondence of scaled total AGN fluxes used in our simulations to physical situations for the remaining red nugget galaxies are shown in Table \ref{tab:scaling_app}, analogous to Table \ref{tab:scaling}. A visual representation of the information in Table \ref{tab:scaling_app} is shown in Figure \ref{fig:straightlines}. Tables \ref{tab:ngc384}, \ref{tab:pgc 11179}, \ref{tab:ugc 2698}, \ref{tab:ngc 1271}, and \ref{tab:mrk 1216} represent danger to species before and after the atmosphere responds to the radiation, for NGC 384, PGC 11179, UGC 2698, NGC 1271, and Mrk 1216, respectively. These are analogous to Tables \ref{tab:m87}, \ref{tab:mw}, and \ref{tab:ngc1277}; they also correspond to the dangerous bolometric AGN fluxes listed for each species and atmospheric composition in Table \ref{tab:fluxes321}.

\begin{table*}[t]
    \centering
    \begin{tabular}{cc||c|c|c|c|c|c|c|} \hline
    \multicolumn{2}{c}{\small Flux ($\rm erg*s^{-1}*cm^{-2}$) =}& $10$ & $10^2$ & $10^3$ & $10^4$ & $10^5$ & $10^6$ & $10^7$ \\ \hline
    \textbf{\footnotesize Galaxy} & \textbf{\footnotesize Edd. Ratio} & \multicolumn{7}{c|}{\textbf{\footnotesize Distance (pc)}} \\ \Xhline{2\arrayrulewidth}

    \multirow{3}{4em}{NGC 384} & $1$ & $6.75 \times 10^3$ & $2.13 \times 10^3$ & $675$ & $213$ & $67.5$ & $21.3$ & $6.75$ \\
    &$0.1$ & $2.13 \times 10^3$ & $675$ & $213$ & $67.5$ & $21.3$ & $6.75$ & $2.13$ \\
    & $0.01$ & $675$ & $213$ & $67.5$ & $21.3$ & $6.75$ & $2.13$ & $0.675$ \\ \hline

    \multirow{3}{4em}{PGC 11179} & $1$ & $1.42 \times 10^4$ & $4.48 \times 10^3$ & $1.42 \times 10^3$ & $448$ & $142$ & $44.8$ & $14.2$ \\
    &$0.1$ & $4.48 \times 10^3$ & $1.42 \times 10^3$ & $448$ & $142$ & $44.8$ & $14.2$ & $4.48$ \\
    & $0.01$ & $1.42 \times 10^3$ & $448$ & $142$ & $44.8$ & $14.2$ & $4.48$ & $1.42$ \\ \hline

    \multirow{3}{4em}{UGC 2698} & $1$ & $1.61 \times 10^4$ & $5.08 \times 10^3$ & $1.61 \times 10^3$ & $508$ & $161$ & $50.8$ & $16.1$ \\
    &$0.1$ & $5.08 \times 10^3$ & $1.61 \times 10^3$ & $508$ & $161$ & $50.8$ & $16.1$ & $5.08$ \\
    & $0.01$ & $1.61 \times 10^3$ & $508$ & $161$ & $50.8$ & $16.1$ & $5.08$ & $1.61$ \\ \hline

    \multirow{3}{4em}{NGC 1271} & $1$ & $1.77 \times 10^4$ & $5.61 \times 10^3$ & $1.77 \times 10^3$ & $561$ & $177$ & $56.1$ & $17.7$ \\
    &$0.1$ & $5.61 \times 10^3$ & $1.77 \times 10^3$ & $561$ & $177$ & $56.1$ & $17.7$ & $5.61$ \\
    & $0.01$ & $1.77 \times 10^3$ & $561$ & $177$ & $56.1$ & $17.7$ & $5.61$ & $1.77$ \\ \hline

    \multirow{3}{4em}{Mrk 1216} & $1$ & $2.27 \times 10^4$ & $7.17 \times 10^3$ & $2.27 \times 10^3$ & $717$ & $227$ & $71.7$ & $22.7$ \\
    & $0.1$ & $7.17 \times 10^3$ & $2.27 \times 10^3$ & $717$ & $227$ & $71.7$ & $22.7$ & $7.17$ \\
    & $0.01$ & $2.27 \times 10^3$ & $717$ & $227$ & $71.7$ & $22.7$ & $7.17$ & $2.27$ \\ \hline

    \end{tabular}
    \caption{\centering Equivalent information to Table \ref{tab:scaling} for the red nugget relic galaxies NGC 384, PGC 11179, UGC 2698, NGC 1271, and Mrk 1216.}
    \label{tab:scaling_app}
\end{table*}

\begin{table*}[t]
    \centering
    \begin{tabular}{ccc||cc|cc|cc|cc|cc|} 
    \hline
    \multicolumn{3}{c}{\textbf{\footnotesize Species}} & \multicolumn{2}{c}{\footnotesize Human} & \multicolumn{2}{c}{\footnotesize \textit{E. Coli}}& \multicolumn{2}{c}{\footnotesize Rat} & \multicolumn{2}{c}{\footnotesize \textit{D. Radiodurans}}\\ \hline
    \textbf{\footnotesize Atmosphere} & \textbf{\footnotesize Time} &\textbf{\footnotesize Edd. Ratio} & \footnotesize Distance (pc) & \footnotesize{Population} & \footnotesize Dist. &  \footnotesize{Pop.} & \footnotesize Dist. &  \footnotesize{Pop.} & \footnotesize Dist. &  \footnotesize{Pop.} \\ \Xhline{2\arrayrulewidth}
    
    \multirow{6}{4em}{Modern} & \multirow{3}{4em}{$t=0$} & $1$ & $181$ & $9.8\%$ &$138$ & $7.8\%$& $31$ & $1.6\%$ & $18$ & $0.6\%$ \\
    &&$0.1$ & $57$ & $3.5\%$ &$44$ & $2.6\%$& $10$ & $<0.2\%$ & $6$ & $<0.2\%$  \\
    && $0.01$ & $18$ & $0.6\%$ &$14$ & $<0.2\%$& $3$ & $<0.2\%$ & $2$ & $<0.2\%$ \\ \cline{2-11}
    
    & \multirow{3}{4em}{$t>36$ d}& $1$ & $63$ & $3.9\%$ &$43$ & $2.5\%$& $8$ & $<0.2\%$ & $<8^*$ & $<0.2\%$ \\
    &&$0.1$ & $20$ & $0.7\%$ &$13$ & $<0.2\%$& $3$ & $<0.2\%$ & $<3^*$ & $<0.2\%$  \\
    && $0.01$ & $6$ & $<0.2\%$ &$4$ & $<0.2\%$& $0.8$ & $<0.2\%$ & $<0.8^*$ & $<0.2\%$ \\ \Xhline{2\arrayrulewidth}

    \multirow{6}{4em}{Proterozoic} & \multirow{3}{4em}{$t=0$} & $1$ & $657$ & $26.3\%$ &$498$ & $21.5\%$& $55$ & $3.3\%$ & $65$ & $4.0\%$ \\
    &&$0.1$ & $206$ & $10.9\%$ &$156$ & $8.7\%$& $17$ & $0.5\%$ & $21$ & $0.8\%$ \\
    && $0.01$ & $65$ & $3.9\%$ & $49$ & $2.9\%$& $5$ & $<0.2\%$ & $7$ & $<0.2\%$ \\ \cline{2-11}

    &\multirow{3}{4em}{$t > 36$ d} & $1$ & $454$ & $20.1\%$ &$330$ & $15.8\%$& $21$ & $0.8\%$ & $9$ & $<0.2\%$ \\
    &&$0.1$ & $143$ & $8.1\%$ &$104$ & $6.2\%$& $7$ & $<0.2\%$ & $3$ & $<0.2\%$ \\
    && $0.01$ & $45$ & $2.7\%$ & $33$ & $1.8\%$& $2$ & $<0.2\%$ & $0.9$ & $<0.2\%$ \\ \Xhline{2\arrayrulewidth}

    \multirow{6}{4em}{Archean} & \multirow{3}{4em}{$t=0$}  & $1$ & $2767$ & $60.4\%$ &$3000$ & $52.5\%$& $299$ & $14.6\%$ & $277$ & $13.8\%$ \\
    &&$0.1$ & $872$ & $31.8\%$ &$661$ & $26.4\%$& $95$ & $5.7\%$ & $87$ & $5.2\%$  \\
    && $0.01$ & $274$ & $13.6\%$ &$207$ & $10.9\%$& $30$ & $1.6\%$ & $27$ & $1.3\%$ \\ \cline{2-11}

    &\multirow{3}{4em}{$t>36$ d} & $1$ & $2377$ & $55.9\%$ &$1810$ & $48.7\%$& $266$ & $13.3\%$ & $241$ & $12.3\%$ \\
    &&$0.1$ & $752$ & $28.9\%$ &$572$ & $23.8\%$& $84$ & $5.1\%$ & $76$ & $4.6\%$  \\
    && $0.01$ & $238$ & $12.2\%$ &$181$ & $9.8\%$& $27$ & $1.3\%$ & $24$ & $1.1\%$ \\ \Xhline{2\arrayrulewidth}

    \end{tabular}
    \caption{\centering \textbf{NGC 384}: Equivalent information to Table \ref{tab:ngc1277}, for the NGC 284 galaxy.}
    \label{tab:ngc384}
\end{table*}

\begin{table*}[t]
    \centering
    \begin{tabular}{ccc||cc|cc|cc|cc|cc|} 
    \hline
    \multicolumn{3}{c}{\textbf{\footnotesize Species}} & \multicolumn{2}{c}{\footnotesize Human} & \multicolumn{2}{c}{\footnotesize \textit{E. Coli}}& \multicolumn{2}{c}{\footnotesize Rat} & \multicolumn{2}{c}{\footnotesize \textit{D. Radiodurans}}\\ \hline
    \textbf{\footnotesize Atmosphere} & \textbf{\footnotesize Time} &\textbf{\footnotesize Edd. Ratio} & \footnotesize Distance (pc) & \footnotesize{Population} & \footnotesize Dist. &  \footnotesize{Pop.} & \footnotesize Dist. &  \footnotesize{Pop.} & \footnotesize Dist. &  \footnotesize{Pop.} \\ \Xhline{2\arrayrulewidth}
    
    \multirow{6}{4em}{Modern} & \multirow{3}{4em}{$t=0$} & $1$ & $374$ & $9.0\%$ &$283$ & $6.6\%$& $64$ & $1.3\%$ & $38$ & $0.6\%$ \\
    &&$0.1$ & $117$ & $2.6\%$ &$89$ & $1.9\%$ & $20$ & $0.2\%$ & $12$ & $<0.04\%$ \\
    && $0.01$ & $37$ & $0.6\%$ &$28$ & $0.4\%$& $6$ & $<0.04\%$ & $4$ & $<0.04\%$ \\ \cline{2-11}
    
    & \multirow{3}{4em}{$t>36$ d}& $1$ & $133$ & $2.9\%$ &$89$ & $1.9\%$& $17$ & $0.1\%$ & $<17^*$ & $<0.1\%^*$ \\
    &&$0.1$ & $42$ & $0.7\%$ &$28$ & $0.4\%$ & $5$ & $<0.04\%$ & $<5^*$ & $<0.04\%$ \\
    && $0.01$ & $13$ & $<0.04\%$ &$9$ & $<0.04\%$& $2$ & $<0.04\%$ & $<2^*$ & $<0.04\%$ \\  \Xhline{2\arrayrulewidth}

    \multirow{6}{4em}{Proterozoic} & \multirow{3}{4em}{$t=0$} & $1$ & $1375$ & $35.4\%$ &$1051$ & $27.8\%$& $114$ & $2.5\%$ & $141$ & $3.1\%$ \\
    &&$0.1$ & $438$ & $10.7\%$ &$334$ & $7.9\%$& $36$ & $0.6\%$ & $44$ & $0.8\%$  \\
    && $0.01$ & $139$ & $3.1\%$ &$106$ & $2.3\%$& $11$ & $<0.04\%$ & $14$ & $0.04\%$ \\ \cline{2-11}

    &\multirow{3}{4em}{$t > 36$ d}& $1$ & $952$ & $25.3\%$ &$693$ & $18.0\%$& $45$ & $0.8\%$ & $20$ & $0.2\%$ \\
    &&$0.1$ & $301$ & $7.1\%$ &$219$ & $5.1\%$& $14$ & $0.04\%$ & $6$ & $<0.04\%$  \\
    && $0.01$ & $95$ & $2.0\%$ &$69$ & $1.4\%$& $4$ & $<0.04\%$ & $2$ & $0.04\%$ \\ \Xhline{2\arrayrulewidth}

    \multirow{6}{4em}{Archean} & \multirow{3}{4em}{$t=0$}  & $1$ & $5669$ & $78.5\%$ &$4349$ & $70.6\%$& $639$ & $16.4\%$ & $588$ & $14.9\%$  \\
    &&$0.1$ & $1813$ & $43.8\%$ &$1387$ & $35.6\%$& $201$ & $4.6\%$ & $186$ & $4.3\%$ \\
    && $0.01$ & $578$ & $14.6\%$ &$441$ & $10.8\%$& $63$ & $1.2\%$ & $59$ & $1.1\%$ \\ \cline{2-11}

    &\multirow{3}{4em}{$t>36$ d} & $1$ & $4987$ & $74.8\%$ &$3797$ & $66.5\%$& $559$ & $14.0\%$ & $505$ & $12.5\%$  \\
    &&$0.1$ & $1577$ & $39.5\%$ &$1201$ & $31.4\%$& $177$ & $4.0\%$ & $160$ & $3.6\%$ \\
    && $0.01$ & $499$ & $12.4\%$ &$380$ & $9.1\%$& $56$ & $1.1\%$ & $50$ & $0.9\%$ \\ \Xhline{2\arrayrulewidth}

    \end{tabular}
    \caption{\centering \textbf{PGC 11179}: Equivalent information to Table \ref{tab:ngc1277}, for the PGC 11179 galaxy.}
    \label{tab:pgc 11179}
\end{table*}

\begin{table*}[t]
    \centering
    \begin{tabular}{ccc||cc|cc|cc|cc|cc|} 
    \hline
    \multicolumn{3}{c}{\textbf{\footnotesize Species}} & \multicolumn{2}{c}{\footnotesize Human} & \multicolumn{2}{c}{\footnotesize \textit{E. Coli}}& \multicolumn{2}{c}{\footnotesize Rat} & \multicolumn{2}{c}{\footnotesize \textit{D. Radiodurans}}\\ \hline
    \textbf{\footnotesize Atmosphere} & \textbf{\footnotesize Time} &\textbf{\footnotesize Edd. Ratio} & \footnotesize Distance (pc) & \footnotesize{Population} & \footnotesize Dist. &  \footnotesize{Pop.} & \footnotesize Dist. &  \footnotesize{Pop.} & \footnotesize Dist. &  \footnotesize{Pop.} \\ \Xhline{2\arrayrulewidth}
    
    \multirow{6}{4em}{Modern} & \multirow{3}{4em}{$t=0$} & $100\%$ & $429$ & $8.5\%$ &$327$ & $5.9\%$& $73$ & $0.9\%$ & $44$ & $0.4\%$ \\
    &&$10\%$ & $136$ & $1.9\%$ &$104$ & $1.4\%$& $23$ & $0.1\%$ & $14$ & $<0.02\%$ \\
    && $1\%$ & $43$ & $0.4\%$ &$33$ & $0.3\%$& $7$ & $<0.02\%$ & $4$ & $<0.02\%$ \\  \cline{2-11}
    
    & \multirow{3}{4em}{$t>36$ d}& $100\%$ & $151$ & $2.2\%$ &$102$ & $1.3\%$& $20$ & $0.09\%$ & $<20^*$ & $<0.09\%^*$ \\
    &&$10\%$ & $48$ & $0.5\%$ &$32$ & $0.2\%$& $6$ & $<0.02\%$ & $<6^*$ & $<0.02\%$ \\
    && $1\%$ & $15$ & $0.03\%$ &$10$ & $<0.02\%$& $2$ & $<0.02\%$ & $<2^*$ & $<0.02\%$ \\ \Xhline{2\arrayrulewidth}

    \multirow{6}{4em}{Proterozoic} & \multirow{3}{4em}{$t=0$}& $1$ & $1569$ & $30.3\%$ &$1189$ & $24.4\%$& $130$ & $1.8\%$ & $157$ & $2.3\%$ \\
    &&$0.1$ & $494$ & $10.1\%$ &$374$ & $7.1\%$& $41$ & $0.4\%$ & $49$ & $0.5\%$ \\
    && $0.01$ & $155$ & $2.2\%$ &$117$ & $1.6\%$& $13$ & $<0.02\%$ & $16$ & $0.04\%$ \\ \cline{2-11}

    &\multirow{3}{4em}{$t > 36$ d}& $1$ & $1081$ & $22.5\%$ &$786$ & $16.7\%$& $51$ & $0.5\%$ & $22$ & $0.1\%$ \\
    &&$0.1$ & $342$ & $6.3\%$ &$249$ & $4.1\%$& $16$ & $0.04\%$ & $7$ & $<0.02\%$ \\
    && $0.01$ & $108$ & $1.4\%$ &$79$ & $0.9\%$& $5$ & $<0.02\%$ & $2$ & $<0.02\%$ \\ \Xhline{2\arrayrulewidth}

    \multirow{6}{4em}{Archean} & \multirow{3}{4em}{$t=0$} & $1$ & $6594$ & $74.1\%$ &$5009$ & $64.0\%$& $705$ & $15.0\%$ & $663$ & $14.1\%$ \\
    &&$0.1$ & $2081$ & $36.9\%$ &$1578$ & $30.4\%$& $226$ & $3.6\%$ & $208$ & $3.3\%$  \\
    && $0.01$ & $655$ & $13.9\%$ &$496$ & $10.1\%$& $72$ & $0.8\%$ & $65$ & $0.7\%$  \\ \cline{2-11}

    &\multirow{3}{4em}{$t>36$ d} & $1$ & $5659$ & $68.6\%$ &$4309$ & $58.4\%$& $634$ & $13.4\%$ & $573$ & $12.0\%$ \\
    &&$0.1$ & $1790$ & $32.3\%$ &$1363$ & $27.2\%$& $201$ & $3.1\%$ & $181$ & $2.7\%$  \\
    && $0.01$ & $566$ & $11.8\%$ &$431$ & $8.5\%$& $63$ & $0.7\%$ & $57$ & $0.6\%$  \\  \Xhline{2\arrayrulewidth}

    \end{tabular}
    \caption{\centering \textbf{UGC 2698}: Equivalent information to Table \ref{tab:ngc1277}, for the UGC 2698 galaxy.}
    \label{tab:ugc 2698}
\end{table*}

\begin{table*}[t]
    \centering
    \begin{tabular}{ccc||cc|cc|cc|cc|cc|} 
    \hline
    \multicolumn{3}{c}{\textbf{\footnotesize Species}} & \multicolumn{2}{c}{\footnotesize Human} & \multicolumn{2}{c}{\footnotesize \textit{E. Coli}}& \multicolumn{2}{c}{\footnotesize Rat} & \multicolumn{2}{c}{\footnotesize \textit{D. Radiodurans}}\\ \hline
    \textbf{\footnotesize Atmosphere} & \textbf{\footnotesize Time} &\textbf{\footnotesize Edd. Ratio} & \footnotesize Distance (pc) & \footnotesize{Population} & \footnotesize Dist. &  \footnotesize{Pop.} & \footnotesize Dist. &  \footnotesize{Pop.} & \footnotesize Dist. &  \footnotesize{Pop.} \\ \Xhline{2\arrayrulewidth}
    
    \multirow{6}{4em}{Modern} & \multirow{3}{4em}{$t=0$} & $1$ & $475$ & $18.6\%$ &$361$ & $14.5\%$& $81$ & $2.9\%$ & $48$ & $1.5\%$ \\
    &&$0.1$ & $150$ & $5.6\%$ &$113$ & $4.2\%$& $26$ & $0.6\%$ & $15$ & $0.1\%$  \\
    && $0.01$ & $47$ & $1.5\%$ &$36$ & $1.0\%$& $8$ & $<0.07\%$ & $5$ & $<0.07\%$  \\  \cline{2-11}
    
    & \multirow{3}{4em}{$t>36$ d} & $1$ & $167$ & $6.3\%$ &$112$ & $4.2\%$& $22$ & $0.4\%$ & $<22^*$ & $<0.4\%^*$ \\
    &&$0.1$ & $53$ & $1.8\%$ &$35$ & $1.0\%$& $7$ & $<0.07\%$ & $<7^*$ & $<0.07\%$  \\
    && $0.01$ & $17$ & $0.2\%$ &$11$ & $<0.07\%$& $2$ & $<0.07\%$ & $<2^*$ & $<0.07\%$  \\ \Xhline{2\arrayrulewidth}

    \multirow{6}{4em}{Proterozoic} & \multirow{3}{4em}{$t=0$} & $1$ & $1690$ & $45.9\%$ &$1288$ & $38.8\%$& $145$ & $5.4\%$ & $175$ & $6.7\%$  \\
    &&$0.1$ & $538$ & $20.8\%$ &$412$ & $16.3\%$&  $46$ & $1.5\%$ & $56$ & $1.9\%$  \\
    && $0.01$ & $172$ & $6.6\%$ &$131$ & $4.9\%$& $14$ & $0.09\%$ & $18$ & $0.2\%$  \\ \cline{2-11}

    &\multirow{3}{4em}{$t > 36$ d} & $1$ & $1193$ & $36.9\%$ &$868$ & $30.2\%$& $56$ & $1.9\%$ & $25$ & $0.6\%$  \\
    &&$0.1$ & $377$ & $15.1\%$ &$275$ & $11.1\%$&  $18$ & $0.2\%$ & $8$ & $<0.07\%$  \\
    && $0.01$ & $119$ & $4.4\%$ &$87$ & $3.2\%$& $6$ & $<0.07\%$ & $2$ & $<0.07\%$  \\ \Xhline{2\arrayrulewidth}

    \multirow{6}{4em}{Archean} & \multirow{3}{4em}{$t=0$} & $1$ & $7173$ & $86.9\%$ &$5422$ & $80.3\%$& $802$ & $28.7\%$ & $722$ & $25.6\%$ \\
    &&$0.1$ & $2249$ & $53.8\%$ &$1698$ & $46.0\%$& $254$ & $10.2\%$ & $231$ & $9.2\%$ \\
    && $0.01$ & $707$ & $26.1\%$ &$543$ & $21.0\%$& $80$ & $2.9\%$ & $73$ & $2.6\%$ \\ \cline{2-11}

    &\multirow{3}{4em}{$t>36$ d} & $1$ & $6250$ & $83.8\%$ &$4759$ & $76.8\%$& $701$ & $26.0\%$ & $633$ & $24.0\%$ \\
    &&$0.1$ & $1976$ & $50.2\%$ &$1505$ & $42.8\%$& $221$ & $8.7\%$ & $200$ & $7.8\%$ \\
    && $0.01$ & $625$ & $23.7\%$ &$476$ & $18.6\%$& $70$ & $2.5\%$ & $63$ & $2.2\%$ \\ \Xhline{2\arrayrulewidth}

    \end{tabular}
    \caption{\centering \textbf{NGC 1271}: Equivalent information to Table \ref{tab:ngc1277}, for the NGC 1271 galaxy.}
    \label{tab:ngc 1271}
\end{table*}

\begin{table*}[t]
    \centering
    \begin{tabular}{ccc||cc|cc|cc|cc|cc|} 
    \hline
    \multicolumn{3}{c}{\textbf{\footnotesize Species}} & \multicolumn{2}{c}{\footnotesize Human} & \multicolumn{2}{c}{\footnotesize \textit{E. Coli}}& \multicolumn{2}{c}{\footnotesize Rat} & \multicolumn{2}{c}{\footnotesize \textit{D. Radiodurans}}\\ \hline
    \textbf{\footnotesize Atmosphere} & \textbf{\footnotesize Time} &\textbf{\footnotesize Edd. Ratio} & \footnotesize Distance (pc) & \footnotesize{Population} & \footnotesize Dist. &  \footnotesize{Pop.} & \footnotesize Dist. &  \footnotesize{Pop.} & \footnotesize Dist. &  \footnotesize{Pop.} \\ \Xhline{2\arrayrulewidth}
    
    \multirow{6}{4em}{Modern} & \multirow{3}{4em}{$t=0$} & $1$ & $609$ & $18.2\%$ & $463$ & $14.3\%$ &$104$ & $2.9\%$ & $61$ & $1.5\%$ \\
    &&$0.1$ & $192$ & $5.7\%$ & $146$ & $4.1\%$ & $33$ & $0.6\%$ & $19$ & $0.2\%$  \\
    && $0.01$ & $61$ & $1.5\%$ &$46$ & $1.1\%$ & $10$ & $<0.05\%$ & $6$ & $<0.05\%$ \\  \cline{2-11}
    
    & \multirow{3}{4em}{$t>36$ d} & $1$ & $213$ & $6.4\%$ & $143$ & $4.1\%$ &$28$ & $0.5\%$ & $<28^*$ & $<0.5\%^*$ \\
    &&$0.1$ & $67$ & $1.7\%$ & $45$ & $1.0\%$ & $9$ & $<0.05\%$ & $<9^*$ & $<0.05\%$  \\
    && $0.01$ & $21$ & $0.3\%$ &$14$ & $0.06\%$ & $3$ & $<0.05\%$ & $<3^*$ & $<0.05\%$ \\ \Xhline{2\arrayrulewidth}

    \multirow{6}{4em}{Proterozoic} & \multirow{3}{4em}{$t=0$} & $1$ & $2193$ & $45.2\%$ &$1658$ & $38.3\%$& $185$ & $5.4\%$ & $220$ & $6.6\%$ \\
    &&$0.1$ & $688$ & $20.1\%$ &$519$ & $15.8\%$& $59$ & $1.4\%$ & $70$ & $1.8\%$ \\
    && $0.01$ &  $215$ & $6.5\%$ &$165$ & $4.7\%$ & $18$ & $0.2\%$ & $22$ & $0.3\%$  \\ \cline{2-11}

    &\multirow{3}{4em}{$t > 36$ d} & $1$ & $1525$ & $36.3\%$ &$1110$ & $29.3\%$& $71$ & $1.8\%$ & $32$ & $0.6\%$ \\
    &&$0.1$ & $482$ & $14.8\%$ &$351$ & $11.0\%$& $23$ & $0.3\%$ & $10$ & $<0.05\%$ \\
    && $0.01$ &  $152$ & $4.3\%$ &$111$ & $3.1\%$ & $7$ & $<0.05\%$ & $3$ & $<0.05\%$  \\ \Xhline{2\arrayrulewidth}

    \multirow{6}{4em}{Archean} & \multirow{3}{4em}{$t=0$} & $1$ & $9264$ & $86.1\%$ &$7018$ & $78.7\%$& $1015$ & $27.5\%$ & $924$ & $25.6\%$    \\
    &&$0.1$ & $2913$ & $52.9\%$ &$2204$ & $45.4\%$& $323$ & $10.1\%$ & $289$ & $9.0\%$ \\
    && $0.01$ & $914$ & $25.4\%$ &$691$ & $20.2\%$& $102$ & $2.8\%$ & $93$ & $2.5\%$  \\  \cline{2-11}

    &\multirow{3}{4em}{$t>36$ d} & $1$ & $7987$ & $82.2\%$ &$6082$ & $74.7\%$& $895$ & $25.0\%$ & $809$ & $23.0\%$    \\
    &&$0.1$ & $2526$ & $49.0\%$ &$1923$ & $41.9\%$& $283$ & $8.8\%$ & $256$ & $7.9\%$ \\
    && $0.01$ & $799$ & $22.8\%$ &$608$ & $18.1\%$& $89$ & $2.4\%$ & $81$ & $2.1\%$  \\ \Xhline{2\arrayrulewidth}

    \end{tabular}
    \caption{\centering \textbf{Mrk 1216}: Equivalent information to Table \ref{tab:ngc1277}, for the Mrk 1216 galaxy.}
    \label{tab:mrk 1216}
\end{table*}

\end{document}